 \documentclass[manuscript]{aastex}

\shorttitle{Estimating Redshifts for Long Gamma-Ray Bursts} \shortauthors{Xiao \& Schaefer }

\begin{document}

\title{Estimating Redshifts for Long Gamma-Ray Bursts}

\author{Limin Xiao and Bradley E. Schaefer}
\affil{Physics and Astronomy, Louisiana State University,
    Baton Rouge, LA, 70803}

\begin{abstract}
The measurement of redshifts for Gamma-Ray Bursts (GRBs) is an important issue for the study of the high redshift universe and cosmology. We
are constructing a program to estimate the redshifts for GRBs from the original \textit{Swift} light curves and spectra, aiming to get
redshifts for the Swift bursts \textit{without} spectroscopic or photometric redshifts. We derive the luminosity indicators from the light
curves and spectra of each burst, including the lag time between low and high photon energy light curves, the variability of the light curve,
the peak energy of the spectrum, the number of peaks in the light curve, and the minimum rise time of the peaks. These luminosity indicators
can each be related directly to the luminosity, and we combine their independent luminosities into one weighted average. Then with our
combined luminosity value, the observed burst peak brightness, and the concordance redshift-distance relation, we can derive the redshift for
each burst. In this paper, we test the accuracy of our method on 107 bursts with known spectroscopic redshift. The reduced $\chi^2$ of our
best redshifts ($z_{best}$) compared with known spectroscopic redshifts ($z_{spec}$) is 0.86, and the average value of
$log_{10}(z_{best}/z_{spec})$ is 0.01, with this indicating that our error bars are good and our estimates are not biased. The RMS scatter of
$log_{10}(z_{best}/z_{spec})$ is 0.26, with a comparison of 0.30 for RMS of $log_{10}(z_{spec})$. We made a selection on bursts with
relatively accurate redshift estimation. The RMS of $log(z_{best}/z_{spec})$ decreases to 0.19, and the RMS scatter of $log_{10}(z_{spec})$
for this subsample is 0.28. For \textit{Swift} bursts measured over a relatively narrow energy band, the uncertainty in determining the peak
energy is one of the main restrictions on our accuracy. Although the accuracy of our $z_{best}$ values are not as good as that of
spectroscopic redshifts, it is very useful for demographic studies, as our sample is nearly complete and the redshifts do not have the severe
selection effects associated with optical spectroscopy.
\end{abstract}


\section{Introduction}
The redshifts have long been an important issue for long-duration Gamma-Ray Bursts (GRBs). Since the first X-ray, optical, and radio
counterparts were discovered in 1997 (Costa et al. 1997; van Paradijs et al. 1997; Frail et al. 1997), GRBs are confirmed to be in galaxies
at cosmological distances. However, until now, only a small fraction of the bursts have their redshifts measured. Even \textit{Swift}, over
its first four years of operation, has only roughly 30$\%$ of its bursts with spectroscopic/photometric redshifts. These redshifts may have
complex selection biases as a function of redshift relating to the difficulty of getting redshifts for faint and distant bursts as well as
the distribution of bands used for spectra and photometry. One illustration of the importance of the selection effects is that the average of
redshift value for \textit{Swift} GRBs have been dropped from $< z >$ = 2.8 (Jakobsson et al. 2006c) and to $< z >$ = 2.1 (Jakobsson 2008)
while the average redshift value of earlier satellites is much lower yet (Bagoly et al. 2009). As the real redshift distribution of GRBs
cannot be changed by that much for the past few years, the only reason for it would be changing selection effects in the spectroscopic
observations.

These systematic biases will greatly affect various analyses. As long-duration GRBs come from the collapse of very massive fast-rotating
stars with very short main-sequence lifetimes (Woosley $\&$ Bloom 2006), their rate density will provide a measurement of the massive star
formation rate of our Universe. If we only deal with bursts with known spectroscopic redshifts, the derived star formation rate may be biased
by the spectroscopic-redshift selection effects. The same problem comes with demographic studies of the GRB luminosity function. If only
bursts with spectroscopic redshifts are used in constructing the luminosity functions, then the selection biases may distort the derived
luminosity function. A third important problem is the recognition of the highest redshift bursts (say, with $z>7$), against which the current
spectroscopic methods may be heavily biased. \textit{Swift} is expected to have $5-10\%$ of its bursts with $z>7$ (Bromm $\&$ Loeb 2006), but
the spectroscopic methods have only identified one such burst GRB090423, and that just a few month ago (Tanvir et al. 2009).

A solution to these problems is to use the luminosity relations that work on luminosity indicators measured directly from the prompt
gamma-ray emission. These luminosity relations are equations that connect a measurable property of the burst itself (the luminosity
indicators) to the burst¡¯s energy ($E_{\gamma}$ or $E_{\gamma,iso}$) or the peak luminosity (L). The observed fluence or peak flux can then
be used (with the inverse-square law) to derive a distance to the burst and (for some fiducial cosmology) the burst redshift. The advantage
of this procedure is that it applies to all long class GRBs, not just to the minority selected to have a measured spectroscopic redshift. The
disadvantage is that the uncertainties on the derived redshifts are much larger than those of the spectroscopic redshifts, and so the method
can be used primarily for statistical or demographic purposes. The situation is similar to the case where photometric redshifts of distant
galaxies in the Hubble Deep Field do not have the accuracy of the spectroscopic redshifts, yet nevertheless these photometric redshifts can
be done for all the galaxies and are the cornerstone of all statistical studies of the field.

The six luminosity relations we use in this article are: $\tau_{lag}$ (spectral lag) - L relation (Norris, Marani, $\&$ Bonnell 2000), V
(Variability) - L relation(Fenimore $\&$ Ramirez-Ruiz 2000), $E_{peak}$ (peak energy of the spectrum) - L relation (Schaefer 2003b),
$E_{peak}$ - $E_{\gamma}$ relation (also called Ghirlanda relation, Ghirlanda, Ghisellini, $\&$ Lazzati (2004)), $\tau_{RT}$ (minimum rise
time) - L relation (Schaefer 2002), and $N_{peak}$ (number of peaks in the light curve) - L relation (Schaefer 2002). It was claimed by
Butler et al. (2007) that the $E_{peak}$ - $E_{\gamma,iso}$ relation (also called Amati relation, Amati et al. (2002)) differs somewhat
between \textit{Swift} and pre-\textit{Swift} data, and hence the relation might be caused by the detection threshold effect of the
instrument instead of the GRB itself. We tested this claim on the relations we are using, as shown in Section 3.

In this article, we will demonstrate a method to use these luminosity relations to estimate the GRBs redshifts. Here, we will only consider
bursts with known spectroscopic redshifts, but our derivation of $z_{best}$ is based on the burst properties, and the effect of the known
spectroscopic redshifts are negligible. This effect is tested later in the article. By applying our method to the bursts with accurate
spectroscopic redshifts measured, we can compare our estimated redshifts with the measured spectroscopic redshifts. This will be the key test
of our methods and the accuracy of our derived redshifts, as well as the applicability of our methods to bursts without spectroscopic
redshift. With confidence in the reliability of the method, we can then apply it to all \textit{Swift} bursts including those with no
spectroscopic redshifts. Also, we will be able to apply this method to future bursts, and provide the community with a rapid notification
about their redshifts. (Xiao $\&$ Schaefer 2009)

\section{Improving the Luminosity Indicators}

In this paper, we adopted five luminosity indicators with six luminosity relations (with two relations for indicator $E_{peak}$: $E_{peak}-L$
relation and $E_{peak}-E_{\gamma, iso}$ relation). The details of the calculations for each of these indicators are described in Schaefer
(2007). In this paper, we constructed the calculation independently, and made some significant improvements in the calculation of
$\tau_{lag}$, $\tau_{RT}$ and $N_{peak}$. The five indicators are listed as below:

 (1)The spectral lag, $\tau_{lag}$, is the delay time between the soft and hard light curves
of a burst. By convention, we use the soft and hard energy bands to be those of BATSE channels 1 and 3, or \textit{Swift} channels 2 and 4,
covering roughly 25-50 keV and 100-350 keV. For $\tau_{lag}$, by shifting the hard and soft light curves of a GRB, and calculating the cross
correlation between them, we are able to get a cross-correlation versus offset plot. The offset corresponding with the peak value of the
cross-correlation is the lag time we need. The calculation is simple and easy for bright bursts, while for those faint ones, since the plot
has significant scatters, the offset with the peak correlation (i.e. $\tau_{lag}$) is hard to evaluate under these noisy conditions. To find
the offset when the cross-correlation achieves its peak value, we need to make a reasonable fit to the peak region of the cross correlation.
As the shape and scatter of the plot varies from burst to burst, we cannot simply fit it with some specific function. If we fit it with a
parabola, then any asymmetry in the cross correlation (as is often seen) will incorrectly shift the peak in the model by an amount depending
on the range of offset included in the fit. And if we fit it with a high-order polynomial function, then the high-order terms will be
unstable as they are trying to follow noise and regions far from the peak. What we did was to fit the cross correlation by polynomials with
different orders (normally from 3 to 9), and choose the one which fits best in the very central region (around the peak) of the curve. Two
examples are shown in Figure 1.

A bootstrap procedure is previously used to calculate the uncertainties on the $\tau_{lag}$ (Norris 2002). In our work, we are using a simple
propagation method. The uncertainties on the cross-correlation amplitude points are calculated by simply evaluating the RMS scatters of these
data points around the fitted curves. We are then able to generate the uncertainties for each of the fitting parameters, and the coefficient
errors. Then the uncertainty for our $\tau_{lag}$ value, $\sigma_{\tau_{lag}}$, is generated by propagation of the uncertainties on the
fitting parameters and the coefficients.

(2) Variability (V) measures whether a light curve is spiky or smooth, and can be obtained by calculating the normalized variance of the
original light curve around the smoothed light curve. The calculation of V is as shown in Schaefer (2007):
\begin{equation}
\label{ } V = <[(C-C_{smooth})^2-\sigma_{C}^2]/C_{smooth,max}^2>
\end{equation}
where C is the count per time bin in the background subtracted light curve, with an uncertainty of $\sigma_{C}$. The time duration of the
burst $T_{up}$ is calculated as the summation of the time with the light curve brighter than 10\% of its peak flux, and $C_{smooth}$ is the
count in the smoothed light curve, with a box-smoothing width to be 30\% of $T_{up}$. $C_{smooth,max}$ is the peak value of $C_{smooth}$.

From equation (1), the $1-\sigma$ uncertainty of V can be propagated from the observational uncertainty of C in each time bin. To get rid of
the cross-correlation effect between C and $C_{smooth}$, for each $C_i$, we calculate it with
\begin{equation}
x_i = C_i-C_{i,smooth} = C_i - \frac{1}{N}(\sum_{j=i-N/2}^{i-1}C_j+\sum_{j=i+1}^{i+N/2}C_j)
\end{equation}
where N is the box smooth bin (i.e. 30\% of $T_{up}$).

If we neglect the uncertainty of $\sigma_C$, the $1-\sigma$ uncertainty of V can be propagated from equation (1) and the uncertainties of
each $x_i$ and $C_{smooth,max}$.

The variability values (V) in Table 1 are about 10 times larger than those in Schaefer (2007). In this article, we are strictly following the
definition and calculation of Equation (1). By checking the details of the calculation, we found that in Schaefer (2007), $C_{max}$ was used
in the denominator instead of $C_{smooth,max}$, which caused the difference. The values of variability are very sensitive to the slightly
change on its calculation function, and this might be one of the reasons for the large scatter of the variability-luminosity relation.

(3) $E_{peak}$ is the photon energy at which the $\nu F_{\nu}$ spectrum is the brightest. By fitting the GRB spectrum with a smoothly broken
power law (Band et al. 1993) as
\begin{eqnarray}
\Phi (E) = \left\{ \begin{array}{ll}
     A~E^{\alpha}~ e^{-(2+\alpha)E/E_{peak}} & \mbox{if $E \le [(\alpha - \beta)/(2+\alpha)]E_{peak}$} \\
     B~E^{\beta} & \mbox{otherwise}
     \end{array}
\right. \
\end{eqnarray}
the peak energy $E_{peak}$ and corresponding parameters $\alpha$ (the asymptotic power law index for photon energies below the break) and
$\beta$ (the power law index for photon energies above the break) can be obtained. Here $\Phi$ is the usual differential photon spectrum
($dN/dE$) as a function of the photon energy (E). A GRB spectrum is also able to be fit with a power-law with exponential cutoff model:
\begin{equation}
 dN/dE = A~E^{\alpha}  e^{-(2+\alpha)E/E_{peak}},
\end{equation}
with $E_{peak}$ and power law index $\alpha$ being the fitting parameters. Here A and B are normalizing constants to indicate the brightness
and constructed to ensure the continuity of the model spectrum.

In this article, we are using $E_{peak}$ values of the bursts from various of sources. All the values we used and the sources are listed in
Table 2.

(4)The minimum rise time in the light curve, $\tau_{RT}$, was proposed for use in a luminosity relation by Schaefer (2002). The minimum rise
time of a burst is taken to be the shortest time over which the light curve rises by half the peak flux of the pulse. In practice, especially
for faint bursts with large Poisson noise, the rate difference between two close bins might be larger than half of the peak flux. As a
result, we have to smooth the light curve before we calculate the rise time. The problem is then that if we smooth it too little, the
apparent fastest rise time might be dominated by the Poisson noise, resulting in a too-small rise time, and if we smooth it too much, the
smoothing effect will dominate, resulting in a rise time near the smoothing time bin.

As the light curves vary greatly among different bursts, there is no specific box smoothing width that can satisfy a majority of bursts.
Instead, we vary the box smoothing width from 0 bins (that is, no binning) up to a relatively large number, say 50 bins, and for each of
these smoothing widths, we generate a smoothed lightcurve from which a minimum rise time can be calculated. Of course, some of these light
curves are over-smoothed and some are under-smoothed. In a minimum rise time versus smoothing width plot, we will have a monotonically rising
curve, as shown in Figure 2. Although the shape of the curve varies amongst bursts, for most of the bursts there will be a region where the
curve appears flat, or with a slightly increasing slope, which we call a plateau. It is easy to explain the existence of the plateau: in this
region, the smoothing is enough that Poisson noise is negligible in determining the minimum rise times, while the smoothing is not so much
that it determines the minimum rise time. On a plot of minimum rise time versus smoothing width, we can identify three regions: a fast rising
region where the Poisson noise dominates, a nearly flat plateau region where we are seeing the real minimum rise time, following by another
rising region for large box smoothing widths where the smoothing is dominating. We can make use of this plateau region by extrapolating it
back to the zero-smoothing case (where the box smoothing width is zero), as shown in Figure 2. The intercept on the y-axis corresponds with
the minimum rise time where the smoothing is effectively zero. With the extrapolation, we are sidestepping the regime where the Poisson noise
dominates. As a result, the value of the intercept is just the minimum rise time we need, not affected by either the Poisson noise effect or
the smoothing effect.

For some of the extremely faint bursts, the Poisson noise dominant region and the smoothing effect dominant region will overlap with each
other, and we are unable to find a plateau in the minimum rise time versus smooth width plot, for which the extrapolation cannot be made.
Thus, our technique does not produce $\tau_{RT}$ values for the faintest bursts.

The uncertainty of the minimum rise time is calculated by simple propagation from the fitting parameters and the uncertainties on each
individual point on the minimum rise time versus smoothing width plot. The uncertainties on each of the individual points are dominated by
the noise on each original data points in the light curve. The noise on the peak flux will affect our criteria, by some factor of
$\sigma_C$/max(C), where C is the rate, and $\sigma_C$ is the uncertainty of the rate. In addition to that, random noise on the start and
stop data point for each possible rise time will affect our determination, i.e. the real rise time between two data points may be
larger/smaller than half of the peak flux, however, with the random noise on the start and stop points, we take it as our rise time, which is
equal to half of the peak flux. This effect is also reflected on our criteria also, by a factor of $\sim2*\sigma_C$/max(C). As a result, for
determine the uncertainties on each of the individual rise time on RT-smoothing width plot, we can change our criteria from 0.5 by a factor
of $2.25*\sigma_C$/max(C), and record how much the resulted minimum rise time values changes. The uncertainties on the fitting parameters and
the minimum rise time value can then be calculated from propagation.

(5)$N_{peak}$ is defined as the number of peaks in the light curve. With $C_{max}$ as the overall maximum of the background-subtracted light
curve, we define a peak to be a local maximum that rises higher than $C_{max}/4$ and is also separated from all other peaks by a local
minimum that is at least $C_{max}/4$ below the lower peak. In principle, $N_{peak}$ is easy to count, either automatically or manually. In
practice, we have the same problem of the Poisson noise and the smoothing factor effect, as what we had in the calculation of $\tau_{RT}$ .
Faint bursts will have their unsmoothed light curves dominated by apparent peaks produced by Poisson noise, resulting in large numbers of
false peaks. A random noise spike can satisfy our definition for a peak if we don¡¯t smooth the light curve, yet if we smooth it too much
there will always be just one peak. Here we adopted the same procedure of calculation as that in the calculation of $\tau_{RT}$ . We vary the
box smooth width from 0 bins to a relatively large number, and calculate the number of peaks for each of the smoothed light curves. As in the
case for $\tau_{RT}$ , we see a fast falling curve (where Poisson noise is contributing spurious peaks), with a plateau, where neither the
Poisson noise effect nor the smoothing effect dominates. By extrapolating the plateau back to the y-axis, we get an $N_{peak}$ value for an
unsmoothed case of the light curve, with the effects of Poisson noise removed.

Each of the luminosity indicators discussed above has one or more corresponding luminosity relations. These relations are $\tau_{lag}-L$,
$V-L$, $E_{peak}-L$, $E_{peak}-E_{\gamma}$ (so called Amati relation) and $E_{peak}-E_{\gamma,iso}$ (so called Ghirlanda relation),
$\tau_{RT}-L$, and $N_{peak}-L$ relation, which are described and explained in details in Schaefer (2007). In our redshift calculation
program, we are not including Amati relation for the following two reasons: First, the Amati relation has been challenged as it returns
ambiguous redshifts (Li 2007), while the $E_{peak}-L$ relation (also the $\tau_{lag}-L$, $V- L$, $\tau_{RT}-L$, and $N_{peak}-L$ relations)
passed the same test (Schaefer \& Collazzi 2007). Second, the physics of the Amati relation is nearly the same as that of Ghirlanda¡¯s
$E_{peak}-E_{\gamma}$ relation, except that Ghirlanda¡¯s $E_{peak}-E_{\gamma}$ relation has a correction for the jet opening angle, making it
much tighter. There is another luminosity relation which relates $E_{peak}$, an effective duration (called $T_{45}$), and the luminosity by
Firmani et al. (2006). However, with more GRBs and $E_{peak}$ data added in, it is realized that this luminosity relation is not making any
improvement on the $E_{peak}-L$ relation (Collazzi \& Schaefer 2007).

\section{Data for 107 Long GRBs and Six Luminosity Relations}

We have 107 long GRBs with their spectroscopic/photometric redshifts measured, ranging from Feb. 28, 1997 (GRB970228) to July 21, 2008
(GRB080721), observed by BATSE, Konus, HETE, and \textit{Swift}. The average redshift for pre-\textit{Swift} bursts is about 1.50 and that of
the \textit{Swift} bursts is 2.15.

The light curves of \textit{Swift} GRBs are generated from the original data published on the legacy ftp
site\footnote{ftp://legacy.gsfc.nasa.gov/swift/data/obs}, and the \textit{Swift} Software ver 2.9 (HEAsoft 6.5). To generate a
background-subtracted light curve of a GRB, we downloaded an event file sw00xxxxxx000.bevshsp\_uf.evt.gz and a mask file
sw00xxxxxx000bcbdq.hk.gz, with xxxxxx be the six digit \textit{Swift} trigger number. By running a task `batbinevt', we can specify the time
interval, energy bins, time bin method, output file name and format on generating the light curve. In our work, we adopted a time interval of
0.064 s with uniform time bins, and four continuous energy bands (15-25 keV, 25-50 keV, 50-100 keV and 100-350 keV). For the calculation of
V, $\tau_{RT}$ and $N_{peak}$, we have been using the light curve over the whole energy range (15-350 keV), and for the calculation of
$\tau_{lag}$ value, we use 25-50 keV and 100-350 keV. The light curves of pre-\textit{Swift} bursts are obtained from our previous work
(Schaefer 2007).

We calculated the $\tau_{lag}$, V, $\tau_{RT}$, and $N_{peak}$ values for each of these bursts, as listed in Table 1. The first column lists
the ID number of GRBs. The second column lists the satellite with the detection of the burst. Column three to six show all the calculated
indicator values, with the name of the indicators shown on the header row. Indicators not measured due to low signal-to-noise ratio of the
burst are represented as `$\ldots$'.

The values of $E_{peak}$ as well as the power law indexes $\alpha$, $\beta$ in Band's smoothly broken power law model (Band et al. 1993) or
$E_{peak}$ and $\alpha$ values in a power law with exponential cutoff model are obtained from various sources, as shown in Table 2. The first
column in Table 2 lists the ID number of the GRBs. The second column lists the satellite with the detection of the burst. Column three to six
are the values of $E_{peak}$ and the power law index $\alpha$ and $\beta$ values, as well as the reference sources of these values. Our jet
break time ($t_{jet}$) values from optical observations and their sources are also listed in Table 2, column seven and eight. Bursts without
measured jet break time are filled by `$\ldots$' also. From the table we see that only 33 of the bursts have their optical $t_{jet}$
reported. Various jet break time in the X-ray afterglows have been reported, however, as there are usually multiple breaks for the X-ray
afterglows, which are not well understood and distinguished for their causes, we are not including any of these reported $t_{jet}$ values
from X-ray detections. Values in square brackets for $\alpha$ and $\beta$ are assumed values for those bursts without exact $\alpha$ and
$\beta$ values measured, which is taken to be the average value of known $\alpha$ and $\beta$ (Schaefer et al. 1994; Krimm, et al. 2009;
Kaneko et al. 2006; Band, et al. 1993). Some uncertainties of $E_{peak}$ and $t_{jet}$ are also quoted in square brackets. These
uncertainties are assumed conservative assumed values, which are normally 10\% of the measured $E_{peak}$ or $t_{jet}$ values. All of the
peak flux (P) and fluence (S) values of pre-\textit{Swift} bursts we use here are as same as what were used in Schaefer (2007), and those of
\textit{Swift} bursts are from the data table on the Swift webpage\footnote{http://swift.gsfc.nasa.gov/docs/swift/archive/grb\_table/}. All
the quoted error bars in Table 1 and Table 2 are converted to $1-\sigma$ level.

The luminosity relations are all expressed as power laws, and we can make a linear fit on the logarithms of the redshift-corrected luminosity
indicators and the logarithms of the burst luminosities. In Figure 3, we display the data and the best fits for each of these luminosity
relations, for the concordance cosmological model (w = -1, and $\Omega_{M}$ = 0.27 in a flat universe). As there are significant
uncertainties in both the horizontal and vertical axes, and some intrinsic scatters independent of both the luminosity and the redshift, we
performed the ordinary least squares without any weighting (Isobe et al. 1990, Schaefer 2007). As both the indicator and luminosity are
caused by some other parameter (like the jet Lorentz factor in particular), and so both indicator and luminosity are correlated. In this
case, we assume that these two variables in the luminosity relations are not directly causative, and the bisector of the two ordinary least
squares (Isobe et al. 1990) has been used. More details of the fitting process are referred to Schaefer (2007). As $E_{peak}$ plays an
important role in the calculation of $P_{bolo}$, the uncertainties on luminosity is then correlated with the uncertainties on $E_{peak}$. In
our work, we ignored that effect, and simply assumed that $P_{bolo}$ and $E_{peak}$ have uncorrelated errors. The best fitting function for
each of the relations are shown in Table 3.

The $V-L$ relation is rather scattered, as we can see from Figure 3 and Table 3. The scatter is larger than one could expected for any linear
relations. As the calculation of V is not well defined, and $V-L$ relation is our most noisy luminosity relations. In this case, we decided
not to include $V-L$ relation in our later calculation for the redshifts, although we listed all the calculated V values for the bursts in
our Tables. As a result, the maximum number of luminosity indicators that we can use is 4, and the maximum number of luminosity relations is
5.

Butler et al. (2007) estimated the Amati's relation ($E_{peak}-E_{\gamma,iso}$) with \textit{Swift} and pre-\textit{Swift} data. They report
an inconsistency in the relations from the two data sets, which they then attributed to differences in threshold between \textit{Swift} and
earlier detectors. Their claimed difference has not been reproduced by other groups (including Cabrera et al. 2007; Schaefer 2007b; Krimm et
al. 2009), while their claimed threshold effects have been found to not significantly affect the observed Amati relation (Schaefer 2007b;
Nava et al. 2008; Ghirlanda et al. 2008; but see Shahmoradi \& Nemiroff 2009). Indeed, their analysis is based on Bayesian priors which
systematically push high-$E_{peak}$ values below $\sim$ 400 keV, as demonstrated by detailed comparisons with Konus, Suzaku, and RHESSI
measures. Nevertheless, in this paper we can perform yet another test to see whether the claimed threshold differences between \textit{Swift}
and pre-\textit{Swift} bursts cause any significant change in the luminosity relations. Butler et al. (2007) paper is including
\textit{Swift} BAT bursts between GRBs 041220 and 070509, 77 of which have spectroscopic redshift measured. While in our analysis on
\textit{Swift} GRBs, we are including all bursts with spectroscopic redshift between 050126 and 080721. With both samples being based on
largely overlapping samples selected in a nearly identical manner, we conclude that the flux limits of the two samples are essentially
identical.

To this end, we have separately fitted the pre-\textit{Swift} and \textit{Swift} data. This test was only done for four luminosity relations
($\tau_{lag}-L$, $V-L$, $E_{peak}-L$, and $\tau_{RT}-L$), with the $E_{peak}-E_{\gamma}$ relation having too few bursts, and the $N_{peak}-L$
relation not being usable for the comparison as a limit. The best fit luminosity relations are given in Table 4. The data and best fit models
are displayed in Figure 4. At first glance, we see that the difference between the two best fit lines are small compared to the scatter in
the data, and a detailed analysis is described below.

We made a F-test for the fitting results for these four relations. First we made a bisector linear fit on the combined data with both
pre-\textit{Swift} and \textit{Swift} bursts, and recorded the $\chi^2$ value of the fit as $\chi_{JOINT}^2$. Then we separate the data to
two sample sets, pre-\textit{Swift} and \textit{Swift}, and made the same bisector linear fit separately on each set of the data. The sum of
the two $\chi^2$ values for the separately fitted lines are recorded as $\chi_{SEPARATE}^2$. Then the F value can be calculated as
\begin{equation}
F = \frac{\chi_{JOINT}^2/(N_{pre}+N_{sw}-2)}{\chi_{SEPARATE}^2/(N_{pre}+N_{sw}-4)}
\end{equation}
where $N_{pre}$ is the number of pre-\textit{Swift} bursts, and $N_{sw}$ is the number of \textit{Swift} bursts. $N_{pre}+N_{sw}-2$ is the
degree of freedom of the fitting on the combined data, and $N_{pre}+N_{sw}-4$ is the degrees of freedom of the fitting on separated data.

These F values for each of the luminosity relations are listed in Table 4. If the pre-\textit{Swift} and \textit{Swift} relations differ much
from each other, the separate fitting have been significantly improved over the fitting on the mixed data, the F value would be much larger
than unity. Otherwise, if there is no significant difference between pre-\textit{Swift} and \textit{Swift} relations, the F value would be
around unity. From Table 4 we see that F is rather close to unity. All of the results shows that the separately fitted result is not
significantly improved over the fitted results on all data mixed together, which means that luminosity relations for pre-\textit{Swift}
luminosity relations and \textit{Swift} luminosity relations do not significantly differ from each other. And by looking at the plots in
Figure 4, we see that (1) the envelope of squares and diamonds are indistinguishable and (2) the pre-\textit{Swift} and \textit{Swift} best
fit lines are close to each other compared to the scatter in the data. And hence, we have no significant evidence that these four luminosity
relations differ for \textit{Swift} bursts.

The $1-\sigma$ range on the normalization difference between \textit{Swift} and pre-\textit{Swift} bursts for all these four luminosity
relations are also listed in Table 4. We can make an analysis with the normalization difference on Amati's relation claimed by Butler et al.
(2007), which is corresponding with a 0.39 difference in log space. By making the comparison between the Butler's factor (0.39) and our
normalization difference, we can exclude Butlers factor at a 2.5 sigma level for $\tau_{lag}-L$ relation, a 3.7 sigma level for $\tau_{RT}-L$
relation, a 1.3 sigma level for $E_{peak}-L$ relation (which is not significant), and we cannot exclude the Butlers factor for $V-L$
relation.

\section{Method for Calculating Redshift}

Our aim in this paper is to test our method of redshift calculation. We applied it to the bursts with known spectroscopic redshifts
($z_{spec}$), and if it works well, we will be able to apply it to all the long GRBs in our future work. Although we are dealing with the
bursts with known $z_{spec}$, these $z_{spec}$ are only involved in the fitting of luminosity relations, and this effect is negligible in our
calculation. Only after we have derived our redshift $z_{best}$ from the luminosity relations will we compare them with the known $z_{spec}$
to test our accuracy.

Below we will describe how our method applies on one Gamma-Ray Burst step by step:

(1) First, we measure each of the luminosity indicators of the burst. The definition and method of calculations have been discussed in
Sections 2 \& 3. The results of the indicators for all bursts in the sample are listed in Table 1 \& 2.

(2) We next derive the luminosity values for each relations from Table 3. A complexity is that the luminosity relations depend on the
redshift of the burst (so as to correct the luminosity indicators back to the burst rest frame), so we have to perform this calculation for
an array of trial redshifts (we take it to be from redshifts of 0 to 20 at intervals of 0.005), and then we will obtain a list of
luminosities (or isotropic energies for Ghirlanda's relation) values depending on the list of trial redshifts for each of the indicators. We
notate each of these calculated luminosities (as a function of redshift $z_{trial}$) for the $i$th relation as $L_{i}(z_{trial})$ (or
$E_{\gamma,i}$ for the Ghirlanda's relation).

(3)With the values of the peak flux P, fluence S, $E_{peak}$ and the power law indexes in the broken power law model $\alpha$ and $\beta$ (or
$\alpha$ from the power law with exponential cutoff model), the bolometric peak flux $P_{bolo}$ and fluence $S_{bolo}$ can be calculated. The
range for `bolometric' is set to be 1 keV to 10000 keV in the GRB rest frame, and the equations for the detailed calculation are referred to
Schaefer (2007). As a result, for each burst, we calculated $P_{bolo}(z_{trial})$ and $S_{bolo}(z_{trial})$ for each trial redshift value
from 0 to 20.

For those bursts with $t_{jet}$ values, the jet opening angle $\theta_{jet}$ (in units of degrees) can be calculated as
\begin{equation}
 \theta_{jet} = 0.161*[\frac{t_{jet}}{1+z_{trial}}]^{3/8}[\frac{n\eta_{\gamma}}{E_{\gamma,iso,52}}]^{1/8},
 \end{equation}
where $t_{jet}$ is the jet break time in the unit of days, n is the density of the circumburst medium in particles per cubic centimeter,
$\eta_{\gamma}$ is the radiative efficiency, and $E_{\gamma,iso,52}$ is the isotropic energy in units of $10^{52}$ erg (Rhoads 1997 \& Sari
et al. 1999). We simply adopt $\eta_{\gamma} = 0.2$ and $n = 3 cm^{-3}$ in equation 6. The beaming factor $F_{beam}$, is then calculated as
\begin{equation}
F_{beam} = 1- \cos\theta_{jet}.
\end{equation}
From above we see that, since both $\theta_{jet}$ and $E_{\gamma,iso}$ are redshift sensitive, our calculated $F_{beam}$ value also varies
between different $z_{trial}$ values.

(4)From all the parameters above, for each of the indicators, a list of the luminosity distances can be calculated as
\begin{equation}
d_{L,i}(z_{trial}) = \sqrt{\frac{L_{i}(z_{trial})}{4\pi * P_{bolo}(z_{trial})}} .
\end{equation}
For Ghirlanda's relation, the list of luminosity distances is calculated as
\begin{equation}
d_{L,i}(z_{trial}) = \sqrt{\frac{E_{\gamma,i}(z_{trial})[1+z_{trial}]}{4\pi F_{beam} S_{bolo}(z_{trial})}} .
\end{equation}
Then from each list of luminosity distance above, the distance modulus can be obtained:
\begin{equation}
\mu_{i}(z_{trial}) = 5 \log [d_{L,i}(z_{trial})] -5,
\end{equation}
with $d_{L,i}(z_{trial})$ expressed in units of parsecs. The uncertainties are propagated strictly following the calculation.

From all above, we are able to get up to four lists of measured distance moduli $\mu_{i}(z_{trial})$ with their $1-\sigma$ uncertainties:
$\mu_{\tau_{lag}} \pm \sigma_{\mu_{\tau_{lag}}}$, $\mu_{E_{peak}}\pm\sigma_{\mu_{E_{peak}}}$, $\mu_{E_{peak}-E_{\gamma}}\pm
\sigma_{\mu_{E_{peak}-E_{\gamma}}}$, and $\mu_{\tau_{RT}} \pm \sigma_{\mu_{\tau_{RT}}}$. As we have asymmetric uncertainties for $E_{peak}$,
we carry the uncertainties on both directions in the calculation, and generated both plus and minus uncertainties for $\mu_{E_{peak}}$ and
$\mu_{E_{peak}-E_{\gamma}}$. $\mu_{N_{peak}}$ is also calculated, as a lower limit on the distance modulus. All of these distance moduli are
a function of the assumed $z_{trial}$ for $0 < z_{trial} < 20$.

(5) Given each trial redshift, we can calculate its distance modulus directly from the cosmological model, $\mu_{cos}(z_{trial})$. Here we
adopt the concordance model, with equation of state for dark energy $p = w \rho c^2 $, $w =-1$, $\Omega_{M} = 0.27$, $\Omega_{\Lambda} = 1 -
\Omega_{M} = 0.73$, and $H_{0} = 70 (km/s)/Mpc$. In this case, the luminosity distance can be expressed as
\begin{equation}
d_L(z_{trial}) = cH_0^{-1} (1+z_{trial}) \int_{0}^{z_{trial}} dz' [(1+z')^3 \Omega_M + \Omega_{\Lambda}]^{-1/2}.
\end{equation}

From the equation above and our list of trial redshifts, a list of luminosity distances $d_{L}(z_{trial})$ will be calculated, and also a
list of distance modulus $\mu_{cos}(z_{trial})$ which equals $5~log[d_{L}(z_{trial})]-5$. The $\mu_{cos}(z_{trial})$ values are only
depending on the trial redshift (running from 0 to 20) and the cosmological model we choose.

(6) For each of the trial redshifts from 0 to 20, we have distance moduli lists of $\mu_{\tau_{lag}}$, $\mu_{E_{peak}}$,
$\mu_{E_{peak}-E_{\gamma}}$,  $\mu_{\tau_{RT}}$ along with their $1-\sigma$ uncertainties as well as $\mu_{cos}$. We can then compare these
$\mu_{i}(z_{trial})$ with $\mu_{cos}(z_{trial})$ in a $\chi ^2$ sense. Thus, $\chi^2_{\tau_{lag}} =
[(\mu_{\tau_{lag}}-\mu_{cos})/\sigma_{\mu_{\tau_{lag}}}]^2$ and so on for the other relations. We get a $\chi_{i}^2$ versus $z_{trial}$ plot,
as shown in the left panel of Figure 5. Then sum over all the $\chi_{i}^2$, we get a $\chi_{total}^2$ ($\chi_{total}^2 = \chi_{\tau_{lag}}
^2+ \chi_{E_{peak}}^2 + \chi_{E_{peak}-E_{\gamma}}^2 + \chi_{\tau_{RT}}^2 + \chi^2_{N_{peak}}$) versus $z_{trial}$ plot , as in the right
panel of Figure 5. Our best redshift ($z_{best}$) corresponds with the minimum $\chi_{total}^2$, where the luminosity relations and the
cosmological model agree with each other best. We are also able to find the uncertainties of our $z_{best}$. By searching through the
$\chi_{total}^2 - z_{trial}$ plot, we can find the redshifts with which the $\chi_{total}^2 = \chi_{total, min}^2 + 1$, which corresponds
with the edges of the $1-\sigma$ range of our redshift. Similarly, the redshifts with $\chi_{total}^2 = \chi_{total, min}^2+ 4$ gives us the
$2-\sigma$ range and that with $\chi_{total}^2 = \chi_{total, min}^2 + 9$ gives the $3-\sigma$ range of our redshift.

\section{Results}

The resulting values of $z_{best}$ and $1-\sigma$ range of $z$ for each burst are shown in Table 5. We have also collected the spectroscopic
redshifts for these bursts (see values and references in Table 5). The last column of Table 5 lists the effective luminosity relations we
used for each burst in our calculation. The number of luminosity indicators used in the redshift calculation for each of the burst is listed
in Table 5. A comparison can be made between our calculated redshifts and their spectroscopic (or photometric) redshifts. The comparison plot
is shown is Figure 6, and below are some of the conclusions after we made the analysis:

(1) Of the total 115 bursts, 8 have only a lower z limit that can be calculated. For the remaining 107 bursts, we took the spectroscopic
redshifts $z_{spec}$ as the model value, and our calculated redshift $z_{best}$ as a measured value, with the $1-\sigma$ uncertainty of
$\sigma_{z_{Plus}}$ and $\sigma_{z_{Minus}}$. Then the $\chi^2$ can be calculated as
\begin{equation}
\chi^2 = \sum_{j=1}^{N_{burst}}{(z_{j, best} - z_{j, spec})^2/\sigma_{j}^2},
\end{equation}
with $\sigma_{j}$ equals to $\sigma_{j,z_{Plus}}$ or $\sigma_{j,z_{Minus}}$, depending on whether our $z_{i, best}$ is smaller or larger than
the $z_{i, spec}$. Each $j$ represents an index number identifying the burst. The number of degrees of freedom in this comparison equals the
number of bursts ($N_{burst}$), so the reduced $\chi^2$ is $\chi^2/N_{burst}$. The reduced $\chi^2$ value is 1.28, which is somewhat larger
than unity. While after excluding one $3-\sigma$ outliers GRB010222, whose contribution to $\chi^2$ is as high as 44, our reduced $\chi^2$ is
equal to 0.86. This is certainly not a significant deviation from unity. So we conclude that the scatter in Figure 6 is consistent with our
quoted error bars being correct.

(2) Of the 107 bursts with their $z_{best}$ calculated, 73 has their $z_{spec}$ falling into the $1-\sigma$ range of our z. The ratio of the
numbers is about $70\%$, which is slightly larger than the ideal case $68.5\%$. And for the 8 bursts with only lower z limit calculated, 6 of
them have their $z_{spec}$ larger than our $1-\sigma$ lower limit of z. This is another way of testing our quoted error bars, and as in the
previous item, we find no significant deviation from the expected results. As such, to a close degree, we see that our derived error bars are
accurate.

(3) We can test to see whether our $z_{best}$ is biased high or low. For this, we calculated the average value of
$log_{10}(z_{best}/z_{spec})$. If our result is unbiased, the average should be zero to within the error bars. We find the average is 0.01.
This demonstrates that our $z_{best}$ is not biased to within the $1\%$ level.

(4) To test the accuracy of our $z_{best}$ comparing with the $z_{spec}$, we calculated the RMS scatter of $log_{10}(z_{best}/z_{spec})$. The
result comes out to be 0.26, while the RMS scatter of $log_{10}(z_{spec})$ is 0.30. In Schaefer (2007) it was pointed out that the accuracy
of the redshift estimation is 26\% (corresponding to a $log_{10}(z_{best}/z_{spec})$ RMS of 0.11), which is better than what we are claiming
here. The reason for the larger RMS scatter is, although we are dealing with the GRBs with known $z_{spec}$, we are not making any use of the
$z_{spec}$ in our whole calculation, and our luminosity L, luminosity distance $d_L$, distance modulus $\mu_{ind}$, as well as the bolometric
flux and fluence $P_{bolo}$ and $S_{bolo}$ are all varying with our trial redshift. This brought us one extra degree of freedom in the
calculation, which caused larger uncertainties in our result. When the known $z_{spec}$ value is used (as for the Hubble diagram work in
Schaefer 2007), the scatter becomes substantially small as compared to our work in this paper. Another reason for the large RMS is because it
is mostly dominated by the very noise bursts (those bursts with low signal to noise ratio, inaccurate measurement of luminosity indicators,
and large error bars for $z_{best}$ results), which are not able to be used for the notification of the redshifts. If we make a selection on
bursts with relatively accurate redshift estimation, say bursts with $\sigma_{z_{minus}/z_{best}}<0.5$ and $\sigma_{z_{plus}/z_{best}}<1$, we
get the RMS of $log(z_{best}/z_{spec})$ of 0.19, which is much smaller, and the RMS scatter of $log_{10}(z_{spec})$ for this subsample is
0.28.

(5)We made the same calculation with pre-\textit{Swift} luminosity relations on calculating \textit{Swift} redshifts and \textit{Swift}
luminosity relations on calculating pre-\textit{Swift} redshifts, for which we call it $z_{best,II}$. Our calculation on $z_{best,II}$ is
totally independent on the spectroscopic redshifts, as for each burst, the luminosity relations used in the calculation are calibrated
independent of the $z_{spec}$ for any individual burst. The comparison between $z_{best}$ and $z_{best,II}$ will also show the difference
between \textit{Swift} and pre-\textit{Swift} luminosity relations. The comparison plot between $z_{spec}$ and $z_{best,II}$ is shown on
figure 7. From the comparison between figure 6 and figure 7 we see that, the scatter and distribution of $z_{best}$ and $z_{best,II}$ do not
differ significantly from each other. Actually, from our calculation, the average value of $log_{10}(z_{best,II}/z_{spec})$ is -0.02, and the
RMS scatter of $log_{10}(z_{best,II}/z_{spec})$ is 0.27, both of which are equal to those of our $z_{best}$ value within error bars. This
result demonstrate that those two sets of $z_{best}$ values do not differ from each other, which means that these two sets of luminosity
relations do not have significant difference. It also tells us that the effect of redshift involved in our calculation (in fitting luminosity
relations) are negligible.

(6)We need to verify whether our result is effective in selecting high redshift bursts. If our predicted redshift is z, the possibility of a
real redshift to be higher than z and lower than z are both $50\%$, which cannot be used as a test. However, by considering our uncertainties
of $z_{best}$, if our predicted redshift is 2z, we can make a test by counting how many of the GRBs are with real redshifts larger than z. As
there are not many bursts with high redshifts, a test is done on a relatively lower redshift region, where most of the GRBs are involved. We
picked up all our GRBs with predicted $z>4$, the total number is 12, and 0 out of 12 have their spectroscopic redshift $z<2$. This result
tells us that our method is actually effective in demographic studies and in picking up high z bursts, if we take into the consideration of
the error bars before we make the prediction.

From all the analysis above, we can conclude that our $z_{best}$ is not biased on average, and our $1-\sigma$ error bars are accurate. We can
claim that our method works well on the bursts with known spectroscopic redshifts, and can be applied to all long GRBs (even without their
$z_{spec}$ measured).

\section{Conclusions}

In this paper, we developed a method to calculate the redshifts for long GRBs, using their light curves and spectra. We applied our method to
bursts with known spectroscopic redshifts, detected by BATSE, HETE, Konus and \textit{Swift}. By comparing our calculated redshifts with
their spectroscopic redshifts, we are able to examine the accuracy of our method.

We compared each of the luminosity relations for pre-\textit{Swift} and \textit{Swift} bursts by making a F-test. With the F values close to
unity, we have significant evidence against any claim that the relations are caused by the detection threshold effects or any other
artificial effects of the instruments.

We compared our results with the spectroscopic redshifts. We find that our $z_{best}$ are not biased (with the average value of
$log_{10}(z_{best}/z_{spec})$ equal to 0.01), and our reported $1-\sigma$ error bars are good (with $\chi_{red}^2 = 0.86$, and $70\%$ of the
$z_{spec}$ fall into the $1-\sigma$ region of $z_{best}$). Our accuracy on the redshifts are not as accurate as those from spectroscopy, yet
nevertheless with a reasonable accuracy for demographical and statistical studies, with the RMS of $log_{10}(z_{best}/z_{spec})$ is 0.26. The
RMS value is about twice as what was found in Schaefer (2007). One of the reason is, in Schaefer (2007), the accuracy is calculated assuming
a known $z_{spec}$, and in this paper, as we are treating the unknown-redshift case, so extra degrees of freedom has been brought in the
calculation, which caused the accuracy to get worse by about a factor of 2. Another reason is that the large RMS is dominated by those faint
and noisy burst, and for a subsample with $\sigma_{z_{minus}/z_{best}}<0.5$ and $\sigma_{z_{plus}/z_{best}}<1$, we get the RMS of
$[log(z_{best}/z_{spec})]$ of 0.19, which is much smaller. As our $z_{best}$ are from the light curves, the spectra and the concordance
cosmological model, it is independent of the spectroscopic redshift. As a result, our method can be applied to all long GRBs.

For \textit{Swift} bursts, as we are measuring over a relatively narrow energy band (15 keV - 350 keV), the uncertainties in the calculation
of peak energy $E_{peak}$ is large, and it becomes one of the main restrictions on our accuracy. With the launch of Fermi, we are hoping to
get bursts with more accurately measured $E_{peak}$ values and light curves covering a broader energy band. We are expecting a substantial
improvement in the accuracy of redshifts for Fermi bursts.

For the next step, we will apply our method to all \textit{Swift} long GRBs, aiming to get a nearly-complete \textit{Swift} GRBs redshift
catalog. Such a catalog will inevitably be incomplete due to bursts with incomplete light curves and bursts too faint for their properties to
be usefully measured. Our resulting redshifts will have an accuracy worse than those obtainable with optical spectroscopy, yet our accuracy
will be good enough for various important statistical studies. As such, our catalog will be used for the demographic studies, without the
detection threshold effect of the spectroscopic redshift measurements. We are also trying to deal with all the future bursts, and to provide
a rapid notification of the redshift on the GCN circular to the community. We are hoping to find some possible high redshift (e.g. $z>7$) and
possible low redshift (e.g. $z<0.3$) GRBs, which will be important in the observation of the high redshift universe and the GRB-SNe
connection study.

\acknowledgements

    This work is supported by NASA under grant NESSF 07Astro07F-0029.

\clearpage

\begin{deluxetable}{llllll}
\tabletypesize{\scriptsize} \tablecaption{Luminosity Indicators. \label{tbl1}} \tablewidth{0pt} \tablehead{ \colhead{$GRB$}   &
\colhead{Satellite} & \colhead{$\tau_{lag}$(sec)}   & \colhead{$V$}   & \colhead{$\tau_{RT}$(sec)} & \colhead{$N_{peak}$} }

\startdata

970228  &   Konus   &       \ldots      &   0.016   $\pm$   0.010   &       \ldots      &   2   \\
970508  &   BATSE   &   0.49    $\pm$   0.02    &   0.018   $\pm$   0.004   &   0.65    $\pm$   0.07    &   1   \\
970828  &   Konus   &       \ldots      &   0.052   $\pm$   0.005   &   0.36    $\pm$   0.14    &   4   \\
971214  &   BATSE   &   0.03    $\pm$   0.05    &   0.048   $\pm$   0.002   &       \ldots      &   2   \\
980703  &   BATSE   &   0.69    $\pm$   0.02    &   0.024   $\pm$   0.001   &   3.00    $\pm$   0.19    &   1   \\
990123  &   BATSE   &   0.07    $\pm$   0.01    &   0.059   $\pm$   0.003   &       \ldots      &   3   \\
990506  &   BATSE   &   0.04    $\pm$   0.01    &   0.337   $\pm$   0.001   &   0.13    $\pm$   0.01    &   12  \\
990510  &   BATSE   &   0.03    $\pm$   0.01    &   0.118   $\pm$   0.001   &   0.13    $\pm$   0.01    &   8   \\
\enddata
\end{deluxetable}

\clearpage

\begin{deluxetable}{llllllllllll}
\tabletypesize{\scriptsize} \tablecaption{Luminosity Indicators. \label{tbl1}} \tablewidth{0pt} \tablehead{ \colhead{$GRB$}   &
\colhead{Satellite}     & \colhead{$E_{peak}$ (keV)\tablenotemark{a}}   & \colhead{$\alpha$\tablenotemark{a}}   &
\colhead{$\beta$\tablenotemark{a}}   & \colhead{Ref.\tablenotemark{b}} & \colhead{$t_{jet}(day)$}& \colhead{Ref.\tablenotemark{b}} }

\startdata
970228  &   Konus   &   $   115 ^   {   +   38  }   _   {   -   38  }   $   &   $   -1.54   \pm 0.08    $   &   $   -2.5    \pm 0.4 $   &   \ldots  &   $       \ldots      $   &   \ldots  \\
970508  &   BATSE   &   $   389 ^   {   +   [40]    }   _   {   -   [40]    }   $   &   $   -1.19   \pm [0.1]   $   &   $   -1.83   \pm [0.4]   $   &   2   &   $   25.00   \pm 5.00    $   &   40  \\
970828  &   Konus   &   $   298 ^   {   +   [30]    }   _   {   -   [30]    }   $   &   $   -0.704  \pm [0.1]   $   &   $   -2.07   \pm [0.4]   $   &   2   &   $   2.20    \pm 0.40    $   &   40  \\
971214  &   BATSE   &   $   190 ^   {   +   [20]    }   _   {   -   [20]    }   $   &   $   -0.78   \pm [0.1]   $   &   $   -2.57   \pm [0.4]   $   &   2   &   $       \ldots      $   &   \ldots  \\
980703  &   BATSE   &   $   254 ^   {   +   [25]    }   _   {   -   [25]    }   $   &   $   -1.31   \pm [0.1]   $   &   $   -2.4    \pm [0.4]   $   &   2   &   $   3.40    \pm 0.50    $   &   40  \\
990123  &   BATSE   &   $   604 ^   {   +   [60]    }   _   {   -   [60]    }   $   &   $   -0.9    \pm [0.1]   $   &   $   -2.48   \pm [0.4]   $   &   2   &   $   2.04    \pm 0.46    $   &   41  \\
990506  &   BATSE   &   $   283 ^   {   +   [30]    }   _   {   -   [30]    }   $   &   $   -1.37   \pm [0.1]   $   &   $   -2.15   \pm [0.4]   $   &   2   &   $       \ldots      $   &   \ldots  \\
990510  &   BATSE   &   $   126 ^   {   +   [10]    }   _   {   -   [10]    }   $   &   $   -1.28   \pm [0.1]   $   &   $   -2.67   \pm [0.4]   $   &   2   &   $   1.60    \pm 0.20    $   &   42  \\
\enddata
\end{deluxetable}

\begin{deluxetable}{llllll}
\tabletypesize{\scriptsize} \tablecaption{Luminosity Relations \label{tbl2}} \tablewidth{0pt} \tablehead{ \colhead{Indicator}   &
\colhead{Relation}   & \colhead{$\sigma_{a}$\tablenotemark{a}}   & \colhead{$\sigma_{b}$\tablenotemark{a}} &
\colhead{$\sigma_{sys}$\tablenotemark{a}} } \startdata

$\tau_{lag}$    &   $\log L = 52.21-0.98 \log[\tau_{lag}(1+z)^{-1}/0.1s]$   &   0.04        &   0.03        &   0.48        \\
$V$ &   $\log L = 51.06 + 1.35 \log[V(1+z)/0.02]$           &   0.09        &   0.09        &   0.87        \\
$E_{peak}$  &   $\log L = 52.10 + 1.88 \log [E_{peak}(1+z)/300 keV]$        &   0.03        &   0.01    &   0.57            \\
$E_{peak}$  &   $\log E_{\gamma} = 50.57 + 1.63 \log [E_{peak}(1+z)/300 keV]$   &   0.09    &   0.03        &   0.16        \\
$\tau_{RT}$ &   $\log L = 52.84 - 1.70 \log [\tau_{RT} (1+z)^{-1}/0.1s]$    &   0.04    &   0.05        &   0.53            \\
$N_{peak}$ & $\log L \ge 50.32+2\log[N_{peak}]~~~~{\rm for}~ N_{peak} \ge 2$ &  \ldots  &   \ldots  &   \ldots      \\
\enddata
\tablenotetext{a}{The $\sigma_{a}$, $\sigma_{b}$ are the $1-\sigma$ uncertainties of the fitting parameters, while \textit{a} is the
intercept on the y-axis, and \textit{b} is the slope of the fitting line. $\sigma_{sys}$ is the systematic uncertainty added in quadrature to
the measurement errors, with which the $\chi_{red}^2$ of the points about the best fit line is unity. }

\end{deluxetable}

\begin{deluxetable}{llllll}
\tabletypesize{\scriptsize} \rotate \tablecaption{Comparison of Luminosity Relations \label{tbl3}} \tablewidth{0pt} \tablehead{
\colhead{Luminosity Relation}   & \colhead{pre-Swift burst}   & \colhead{Swift burst}     & \colhead{F}   &
\colhead{$\sigma_{norm,diff}$\tablenotemark{a}} } \startdata

$\tau_{lag}$-L  &   $\log L = 52.14-0.95 \log[\tau_{lag}(1+z)^{-1}/0.1s]$   &   $\log L = 52.26-1.00 \log[\tau_{lag}(1+z)^{-1}/0.1s]$ & 0.98   &    0.11 \\
Variability-L   &   $\log L = 51.38 + 1.24 \log[V(1+z)/0.02]$           &   $\log L = 50.92 + 1.34 \log[V(1+z)/0.02]$   & 1.04 &   0.09 \\
$E_{peak}$-L    &   $\log L = 51.42 + 1.71 \log [E_{peak}(1+z)/300 keV]$        &   $\log L = 51.13 + 1.88 \log [E_{peak}(1+z)/300 keV]$ & 1.02 & 0.08 \\
$\tau_{RT}$-L   &   $\log L = 52.87 - 1.44 \log [\tau_{RT} (1+z)^{-1}/0.1s]$    &   $\log L = 52.81 - 1.85 \log [\tau_{RT} (1+z)^{-1}/0.1s]$ & 1.00  & 0.18 \\

\enddata
\tablenotetext{a}{The $\sigma_{norm,diff}$ is the $1-\sigma$ range of the difference of normalization values between \textit{Swift} and
pre-\textit{Swift} luminosity relations. It is calculated by $\sqrt{\sigma^2_{norm,preSwift}+\sigma^2_{norm,Swift}}$.}
\end{deluxetable}

\clearpage

\begin{deluxetable}{llllllllll}
\tabletypesize{\scriptsize} \tablecaption{Our Redshifts and Spectroscopic Redshifts. \label{tbl4}} \tablewidth{0pt} \tablehead{
\colhead{$GRB$}   & \colhead{Experiment} & \colhead{$z_{spec}$}   & \colhead{Ref.\tablenotemark{a}}  & \colhead{$z_{best}$}   &
\colhead{$z_{min}$}   & \colhead{$z_{max}$}  & \colhead{$N_{rel}$} } \startdata
970228  &   Konus   &   0.84    &   1   &   0.38    &   0.18    &   0.86    &   2   \\
970508  &   BATSE   &   0.84    &   2   &   1.51    &   1.11    &   2.09    &   4   \\
970828  &   Konus   &   0.96    &   3   &   1.10    &   0.73    &   1.83    &   4   \\
971214  &   BATSE   &   3.42    &   4   &   5.21    &   2.00    &   20  &   3   \\
980703  &   BATSE   &   0.97    &   5   &   0.84    &   0.66    &   1.11    &   4   \\
990123  &   BATSE   &   1.61    &   6   &   0.91    &   0.67    &   1.27    &   4   \\
990506  &   BATSE   &   1.31    &   7   &   1.16    &   0.77    &   1.84    &   4   \\
990510  &   BATSE   &   1.62    &   8   &   2.47    &   1.69    &   3.66    &   5   \\
\enddata
\end{deluxetable}

\clearpage

\begin{figure}
\epsscale{1.10} \plottwo{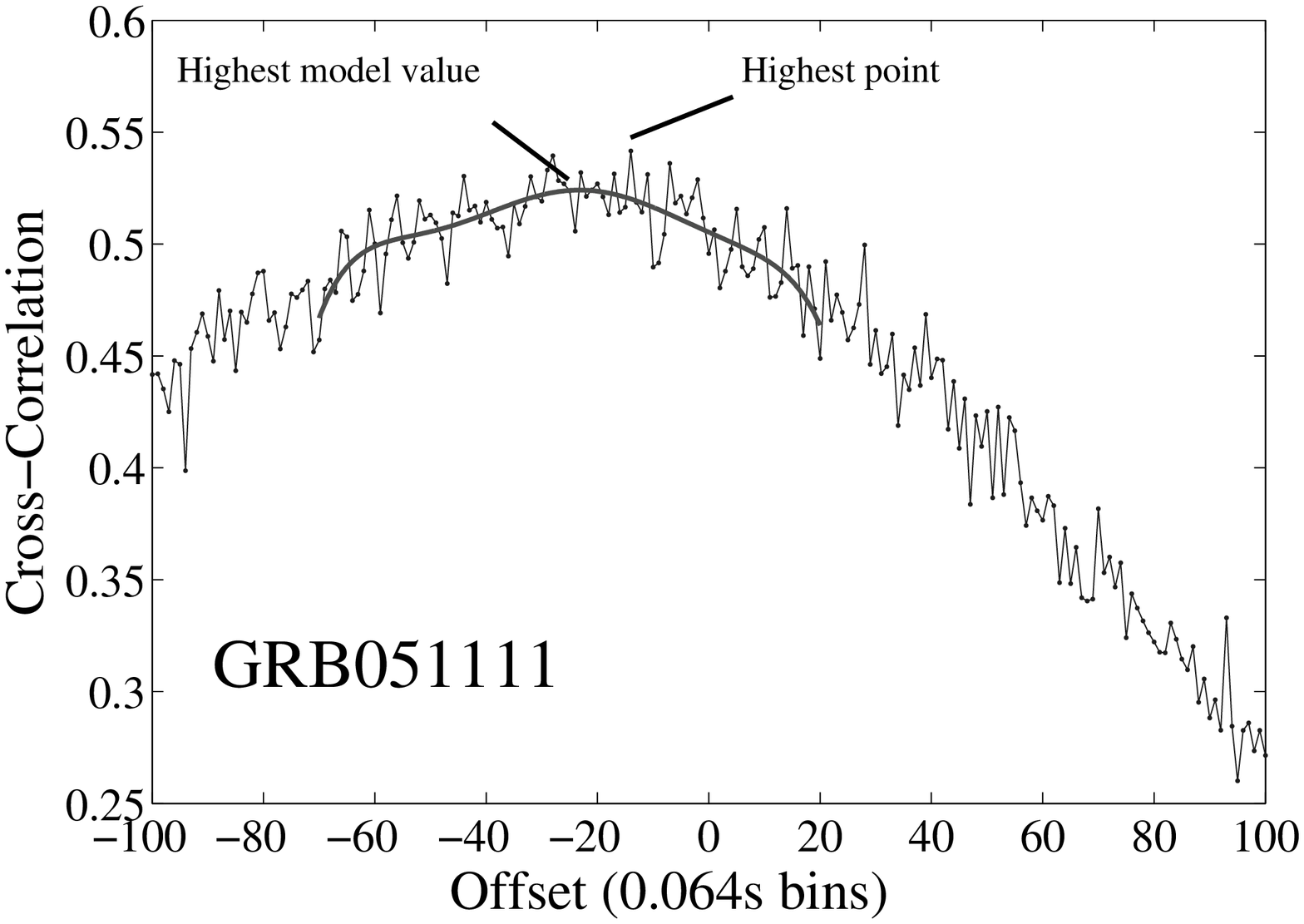}{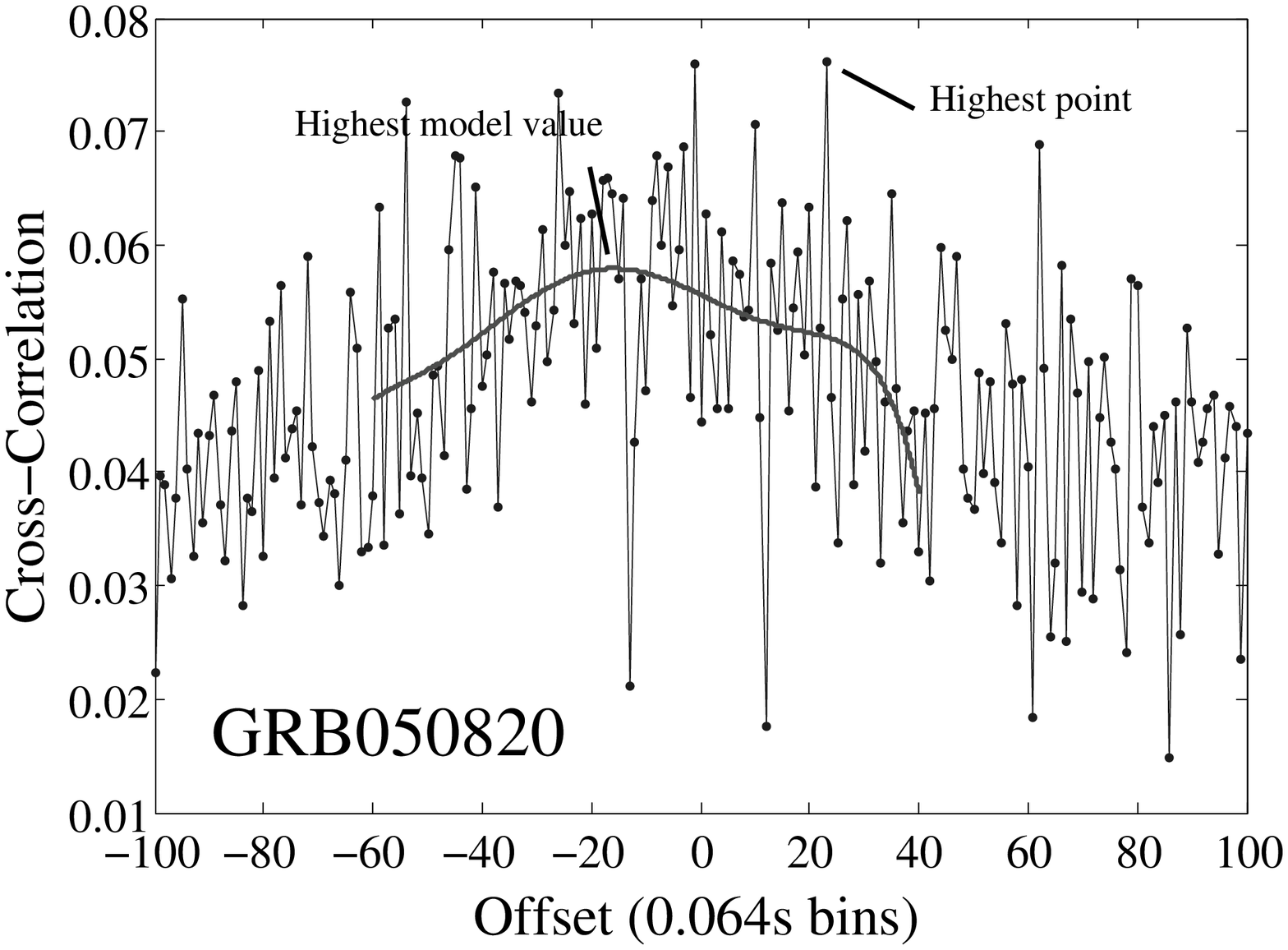} \caption{$\tau_{lag}$ calculation -- Cross-Correlation versus Offset plots. The left panel is the
plot for GRB051111, which is a moderately bright burst without much noise in the light curve. But we can see that although the curve is
relatively smooth, the peak region is quite 'broad' that we still need an accurate fitting of the curve, especially for the central region of
the curve. Had we simply taken the offset of the highest measured cross-correlation, we would have the peak at -14 bins (with $\tau_{lag}$
equal to 0.90 seconds), with this value being determined by noise in the cross correlation. Instead, we fitted a 7-order polynomial (smooth
curve) and found a peak at -23.1 bins (with $\tau_{lag}$ equal to 1.48 seconds), and this model fit is avoiding most of the random noise in
the cross correlation. Note, for our choice of bands, the lag will be negative in bins but expressed as positive in time. The right panel is
for GRB050820, which is a noisy burst. There is much scatter in the cross-correlation versus offset plot, from which we cannot determine the
position of the peak without making a fit around the peak region. In this case, the highest cross correlation is at +23 bins (corresponding
to a negative lag of -1.47 seconds), with the model fit (a 7-order polynomial also) providing a more reasonable peak at -15.7 bins
($\tau_{lag}$ = 1.00 seconds). }
\end{figure}

\begin{figure}
\epsscale{0.80} \plotone{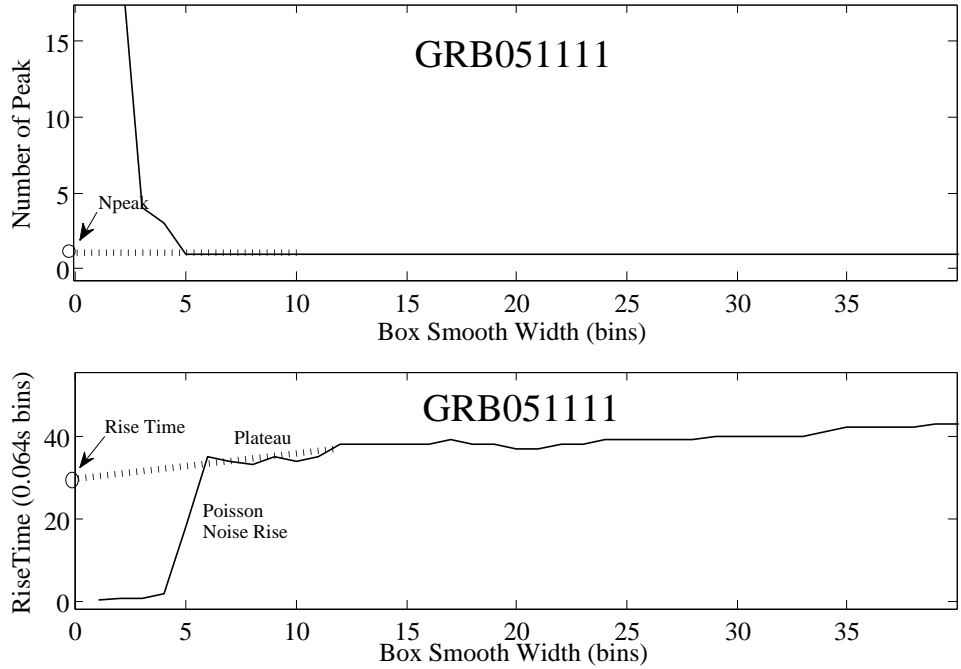} \caption{ The calculation of $\tau_{RT}$ and $N_{peak}$. We can see an apparent 'plateau' in the lower plot,
for box smoothing widths of more than 5. By simply extrapolating the plateau back to the y-axis (see dashed line), we get a minimum rise time
(see circle at intercept on y-axis) which is appropriate for zero smoothing, and which is not affected by the Poisson noise. }
\end{figure}

\begin{figure}
\epsscale{1.05} \centering \plottwo{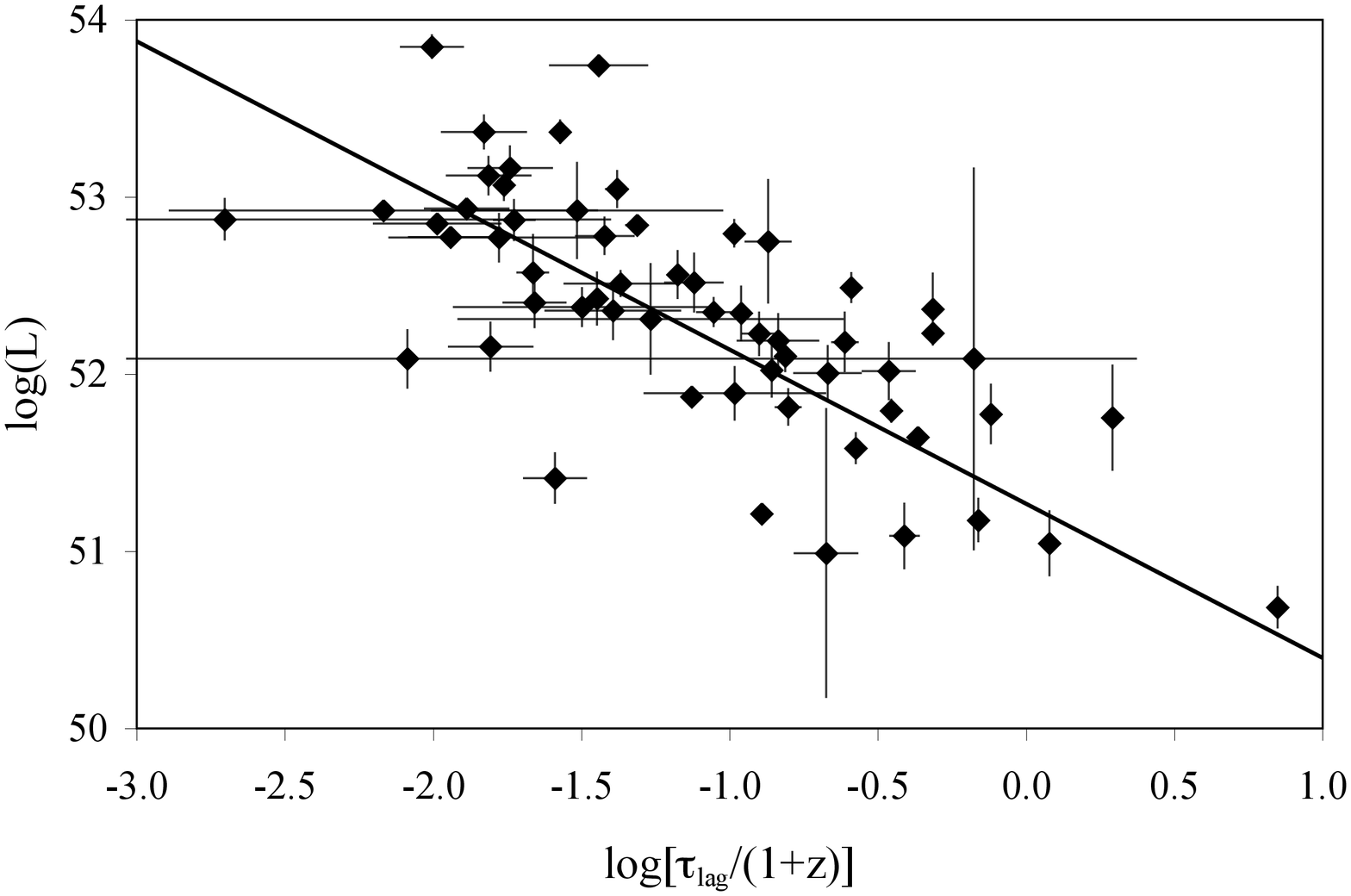}{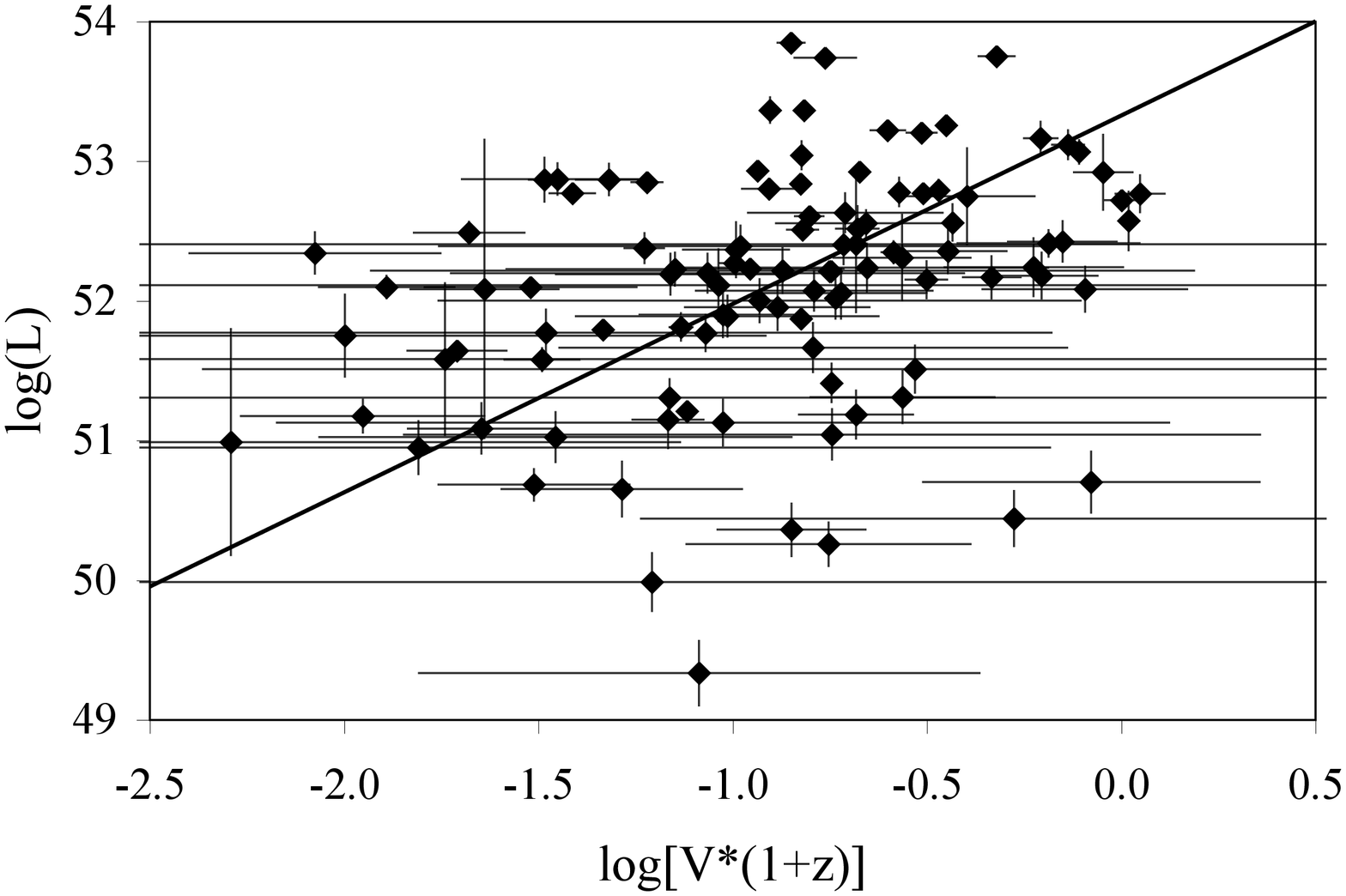} \plottwo{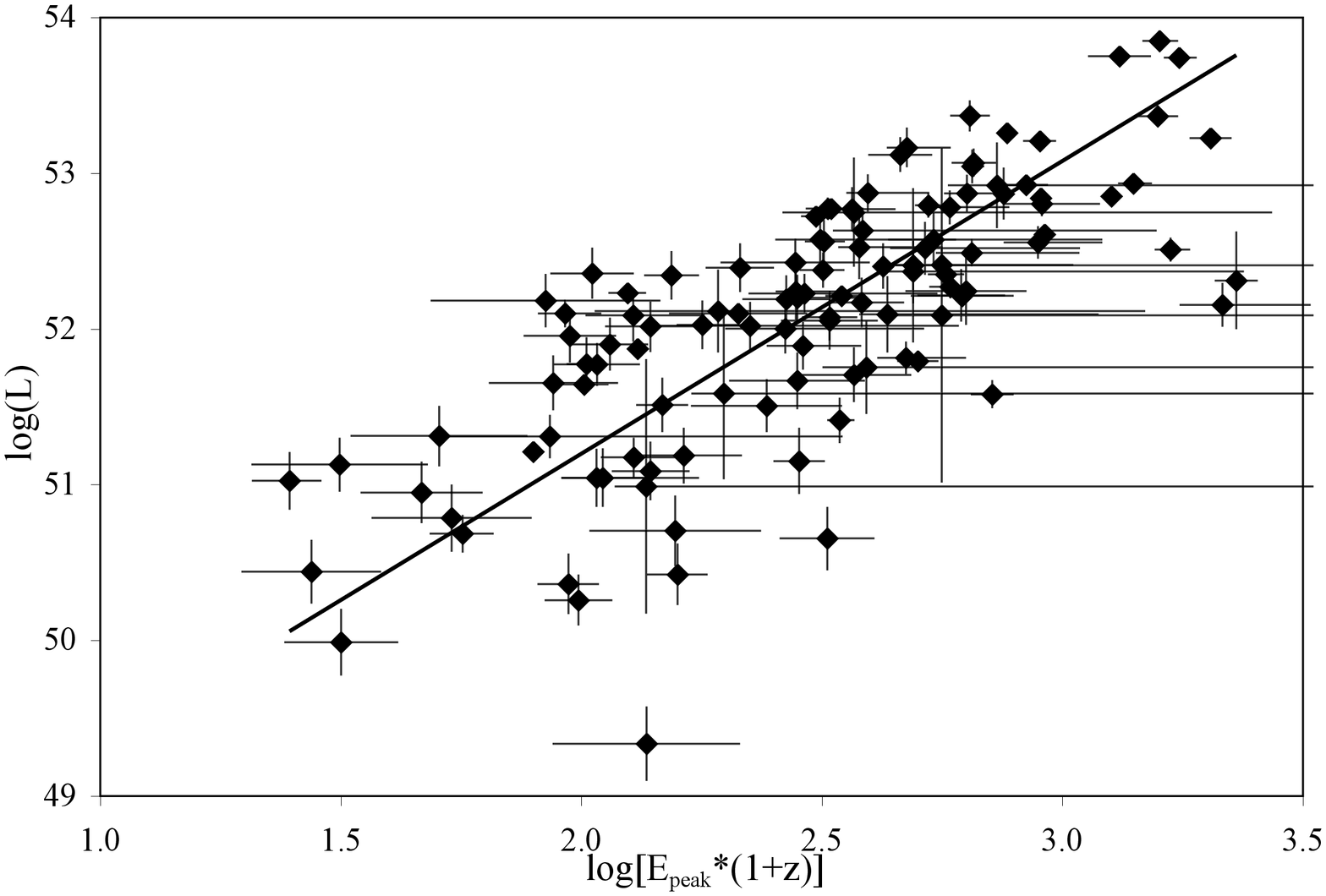}{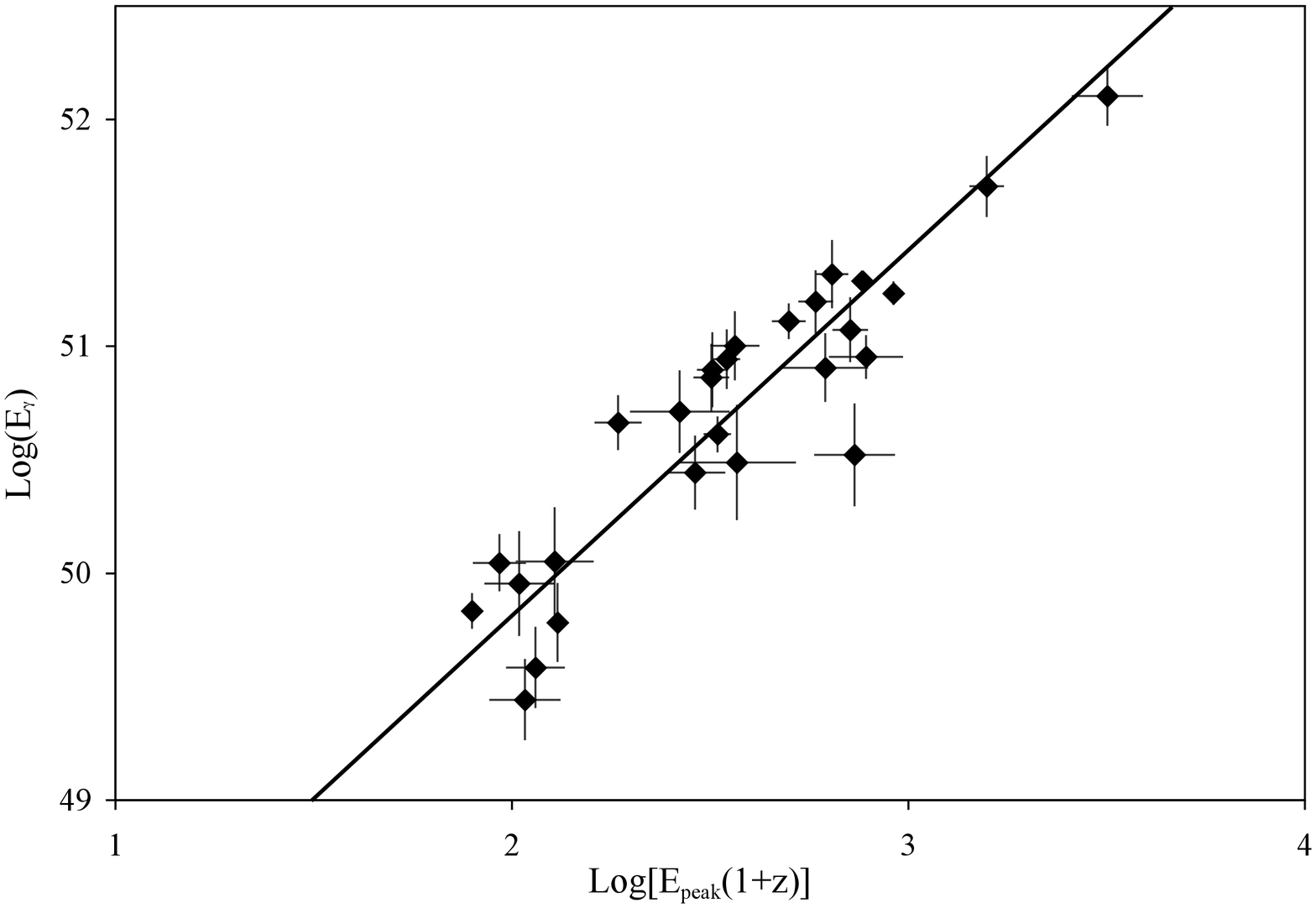} \plottwo{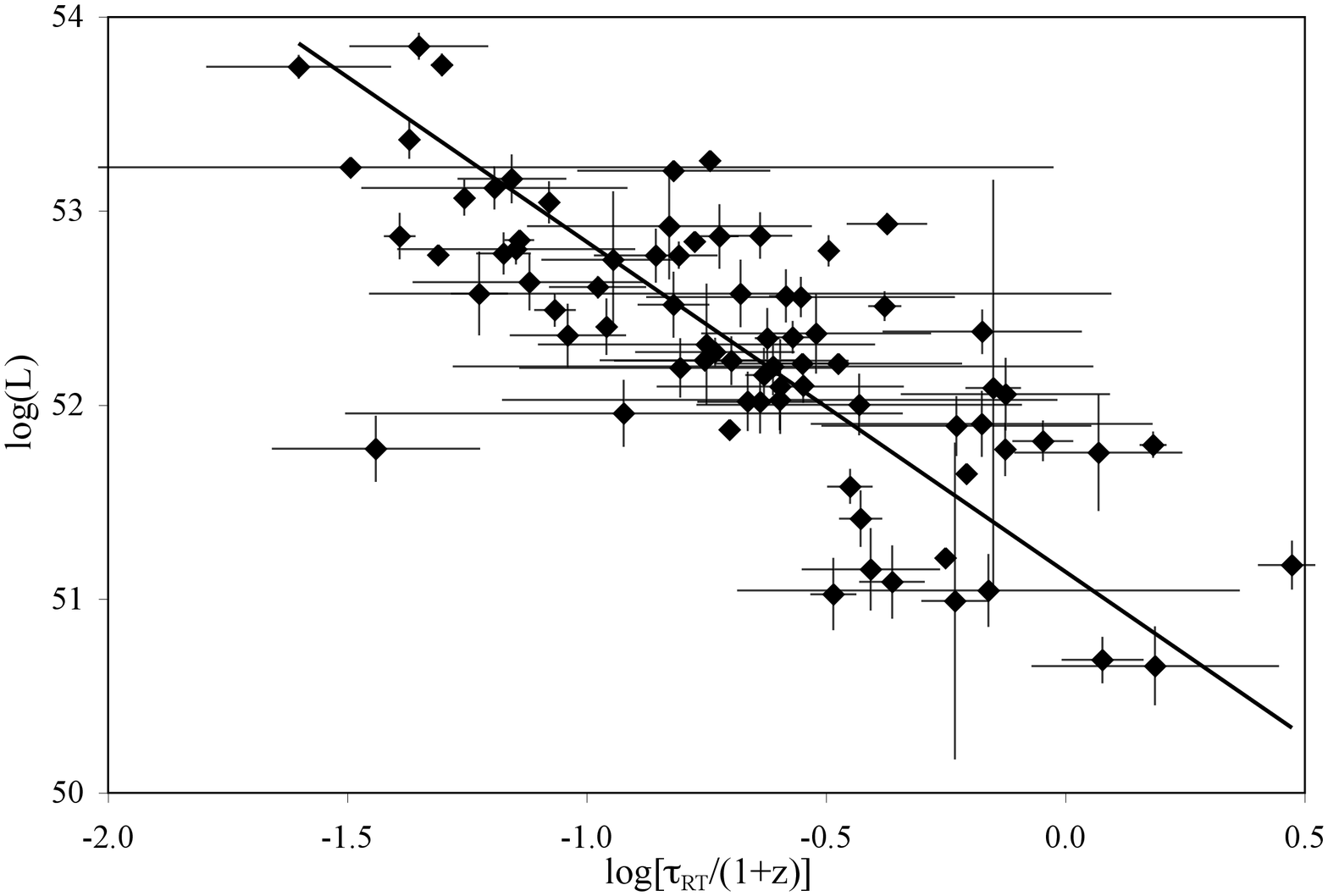} {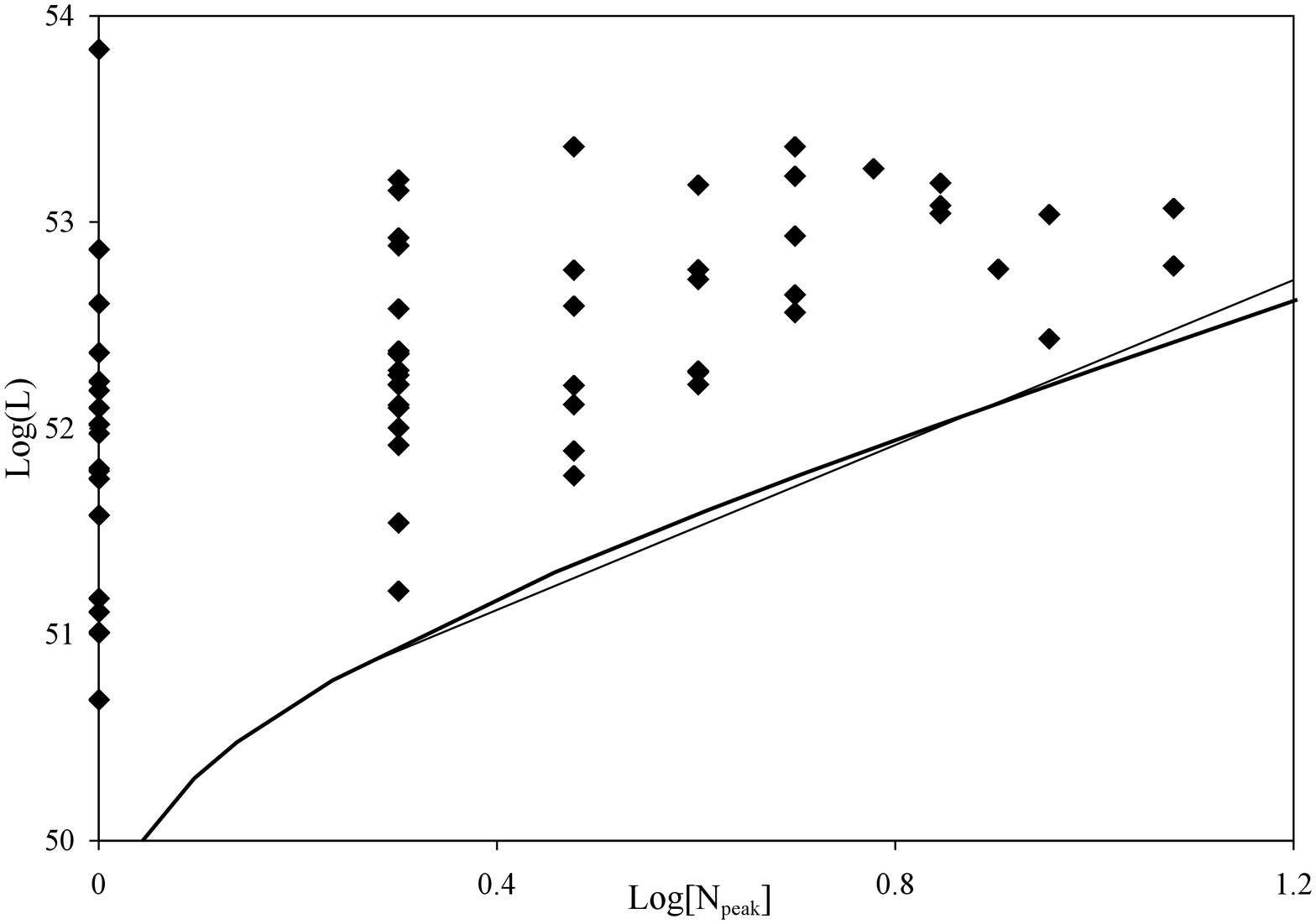} \caption{ Luminosity relations
for long GRBs. $\tau_{lag}$, $V$, $\tau_{RT}$ and $N_{peak}$ values are generated from light curves of the GRBs, while $E_{peak}$ values are
generated from the spectra. The $1-\sigma$ measurement uncertainties are used as the error bar. The best fit relation of the six is the
$E_{peak}-E_{\gamma}$ relation (Ghirlanda's relation), while the most scattered one is the $V-L$ relation. However, with the restrictions on
the measurements of $E_{peak}$ (especially for Swift bursts) and jet break time, not many bursts can be included in the Ghirlanda's relation.
The $N_{peak}-L$ relation provides only a lower limit on the luminosity, but this is a strong limit never violated, for example where a
10-peak burst always has $L > 10^{52.32}$ ergs.}

\end{figure}

\clearpage

\begin{figure}
\epsscale{1.10} \plottwo{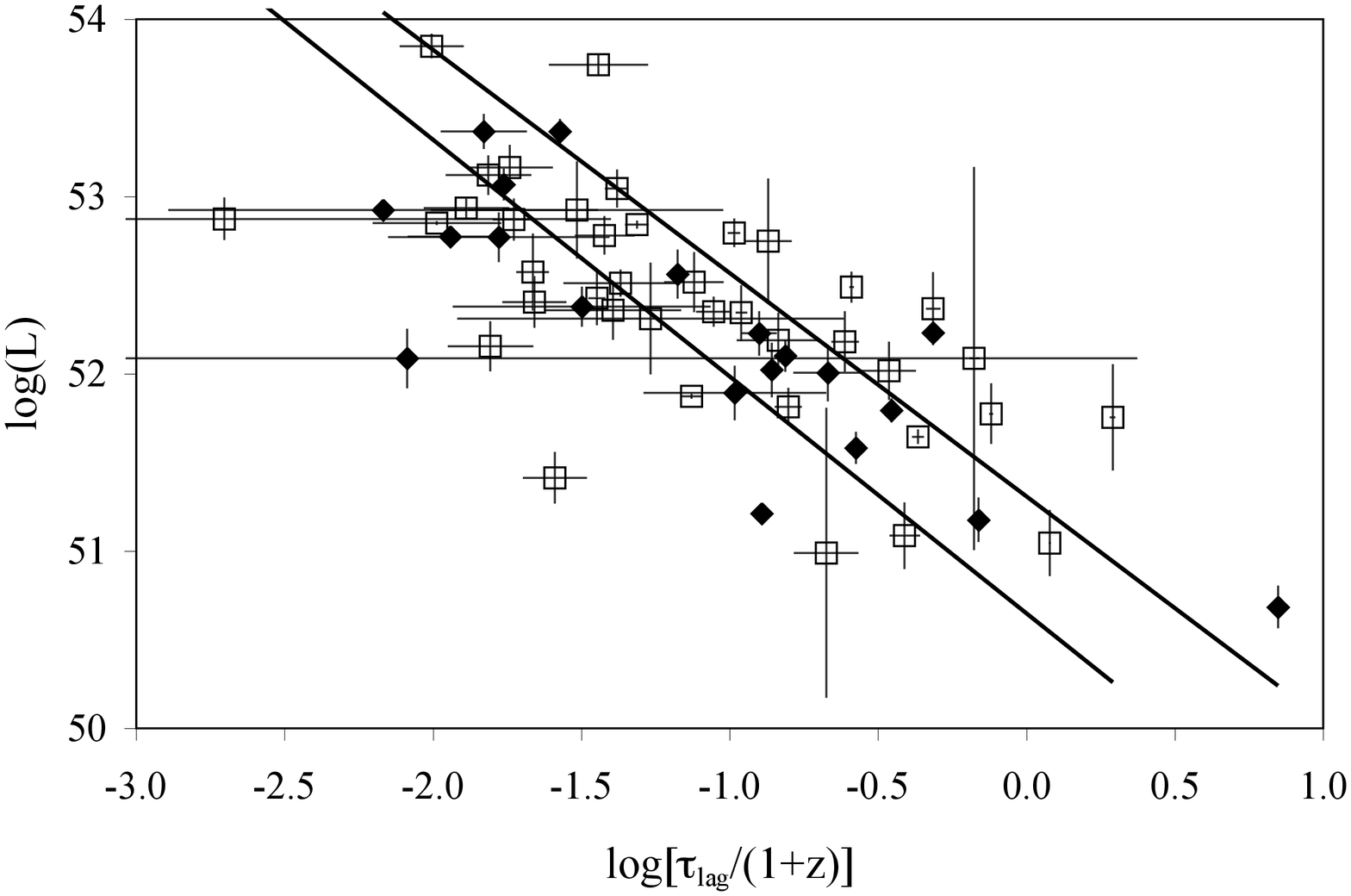}{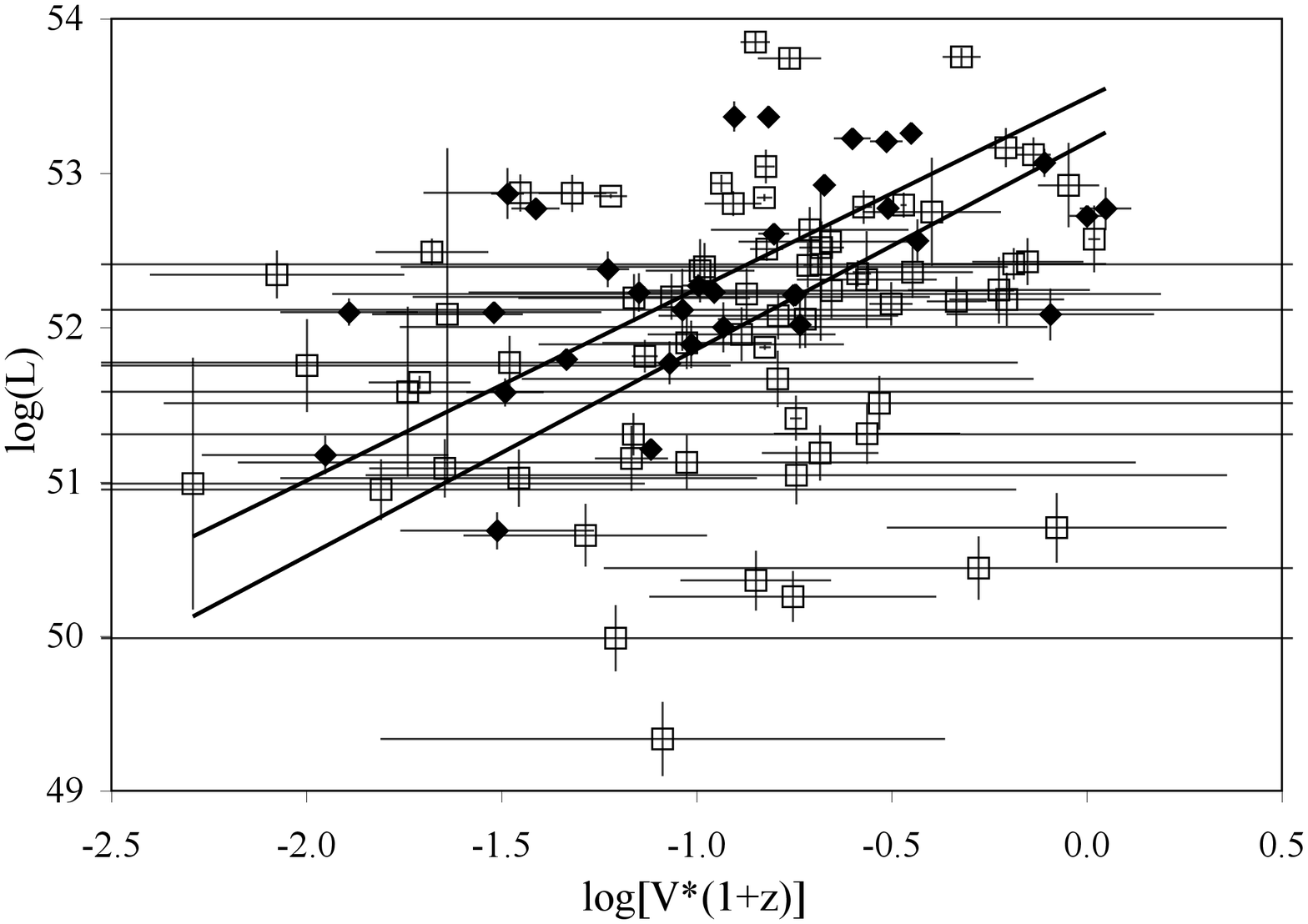} \plottwo{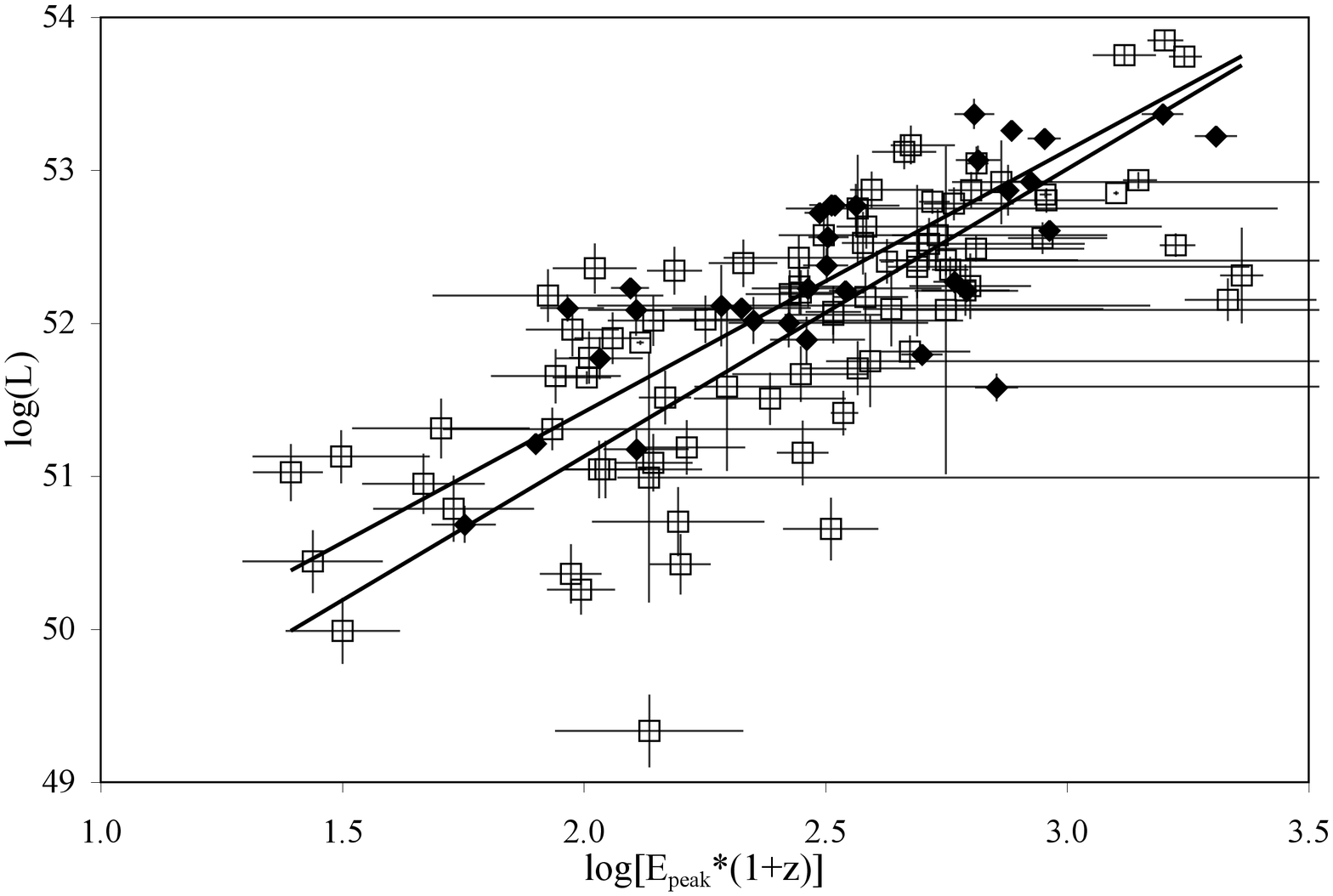}{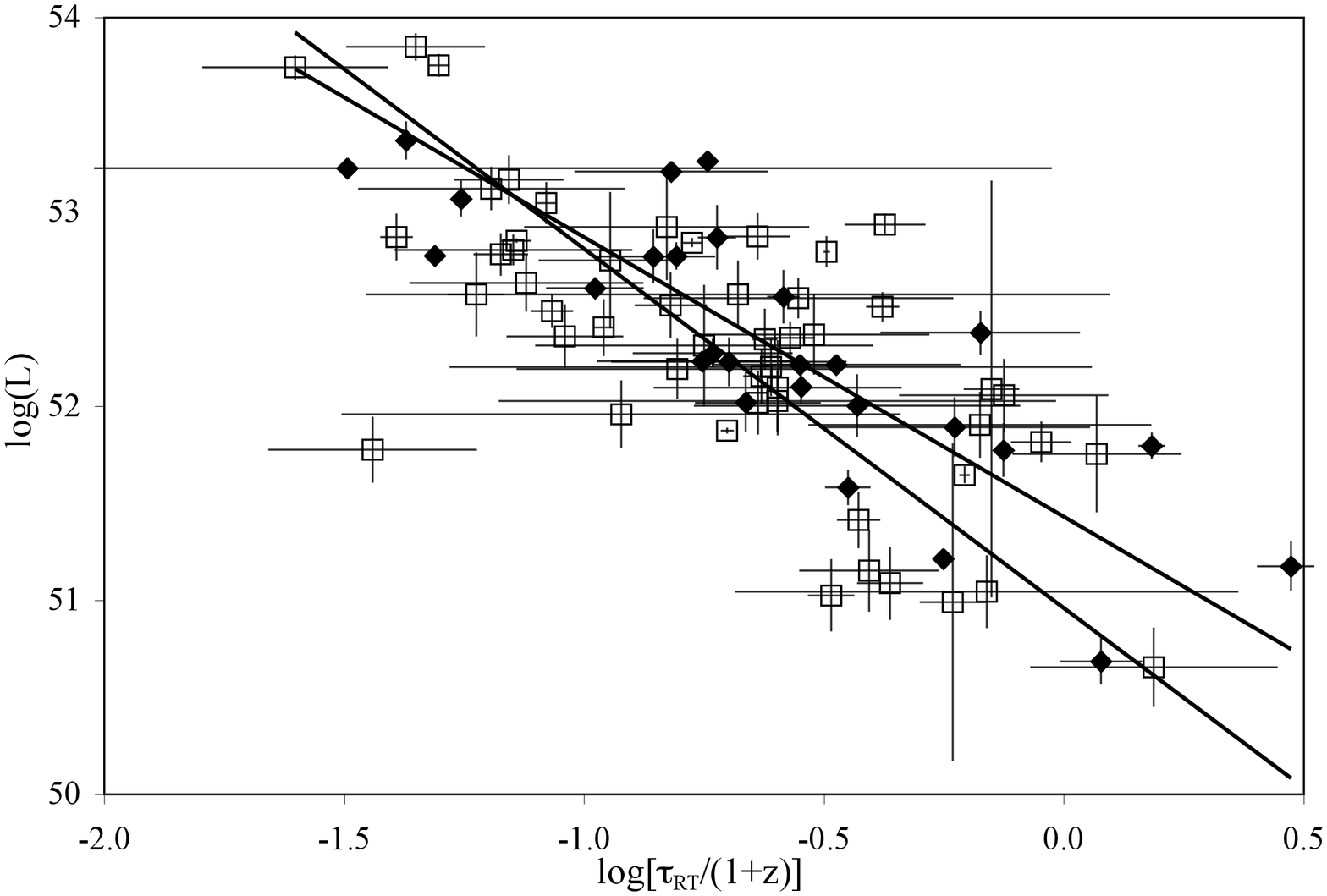} \caption{ Comparison of the luminosity relations for pre-Swift and
Swift bursts. The filled diamonds are for pre-Swift bursts, while the open squares are for Swift bursts. From the plot we can tell that the
luminosity relations are identical between pre-Swift and Swift bursts to within the usual uncertainties. Ghirlanda's relation is not compared
because the sample of bursts with both $E_{peak}$ and jet break time is too small. }

\end{figure}

\clearpage

\begin{figure}
\epsscale{1.10} \plottwo{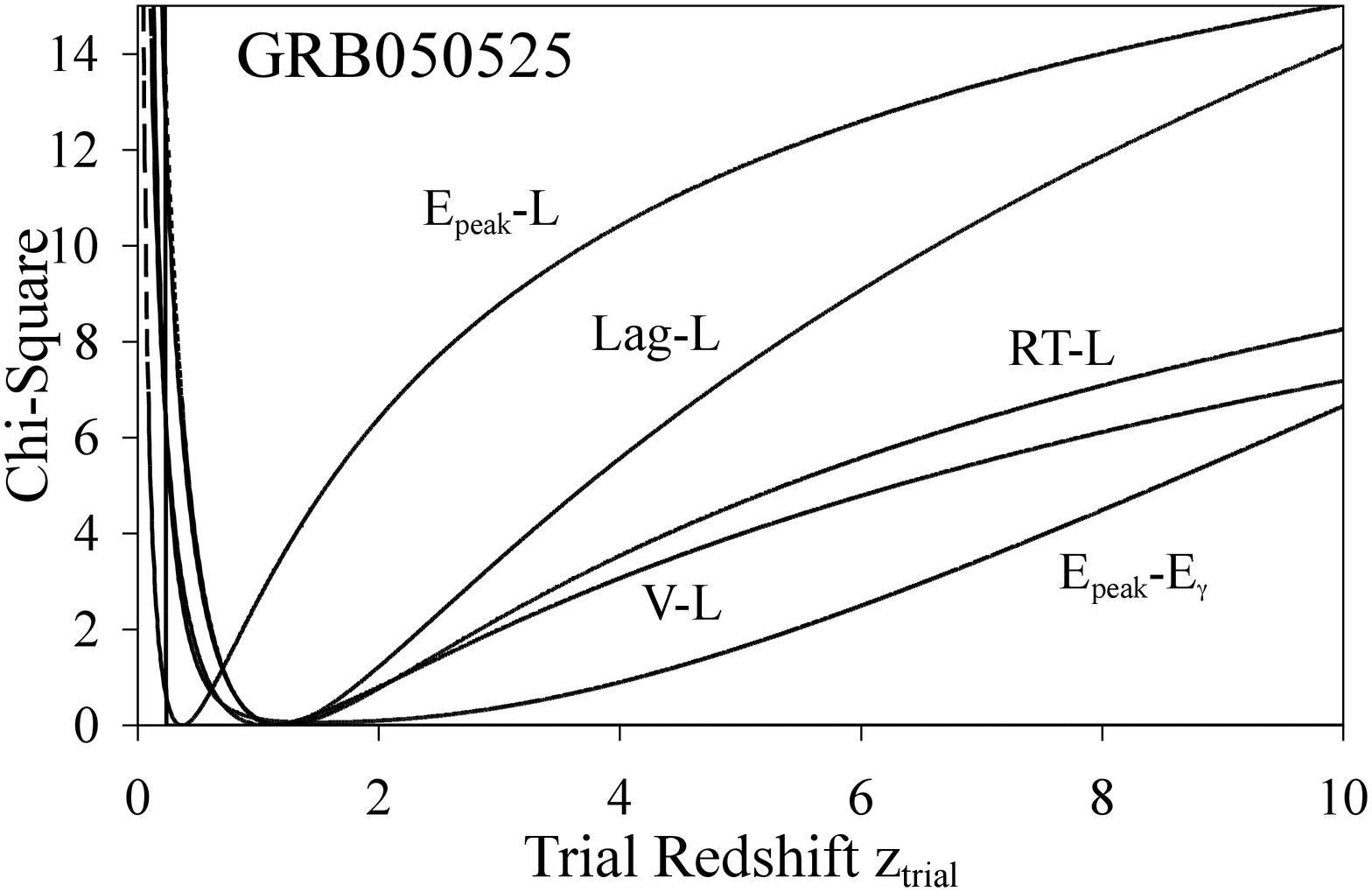}{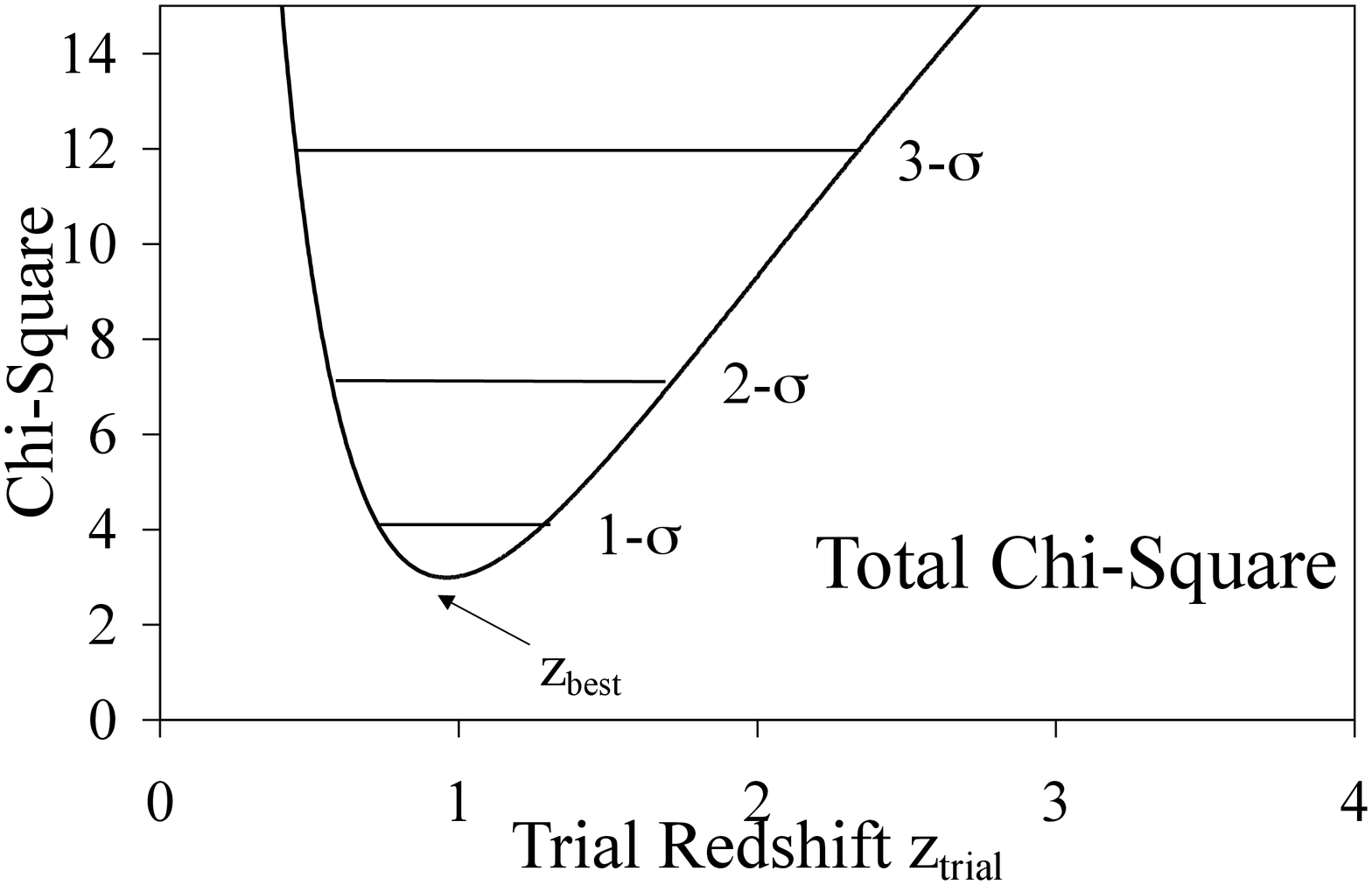} \caption{ One example of the $\chi^{2}$ plot in our calculation. Taking the $\mu_{cos}$ as model
value and each of the $\mu_{i}$ as observed value, we can have the $\chi^{2}$ for each of the indicators, as shown in the left panel. Each of
the luminosity relations corresponding with these curves are labeled just beside the lines, and the vertical line on the left is the
luminosity lower limit given by the $N_{peak}-L$ relation. We sum up all the included $\chi^{2}$s to get a total $\chi^{2}$ plot, as shown in
the right panel. Then the minimum $\chi^{2}$ corresponds with our $z_{best}$, and $\chi^2 = \chi^{2}_{min}$ + 1, 4, 9 gives us the
1,2,3-$\sigma$ range of our estimated redshifts. }

\end{figure}

\clearpage

\begin{figure}
\epsscale{0.80} \plotone{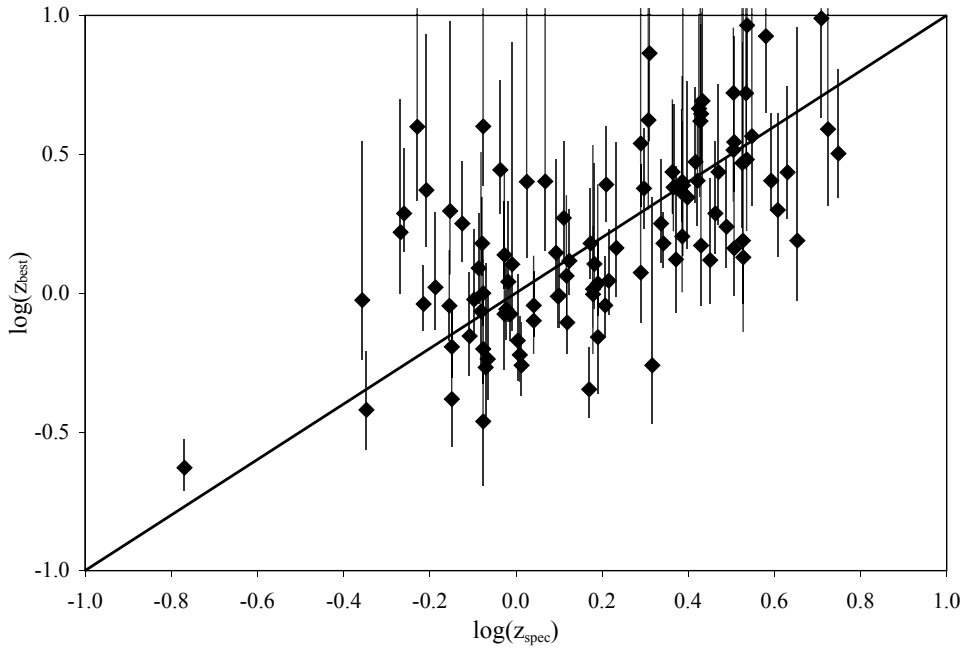} \caption{Comparison of our redshifts with spectroscopic redshifts. The diamonds are our $z_{best}$ and the
error bars gives the 1-$\sigma$ redshift range. We can see that 70\% of the spectroscopic redshifts fall in our 1-$\sigma$ range, and our
$z_{best}$ scatter uniformly around the spectroscopic redshift (with $\langle$$log(z_{best}/z_{spec})$$\rangle$ = 0.01). This demonstrates
that our method has accurate error bars and negligible biases. }

\end{figure}

\begin{figure}
\epsscale{0.80} \plotone{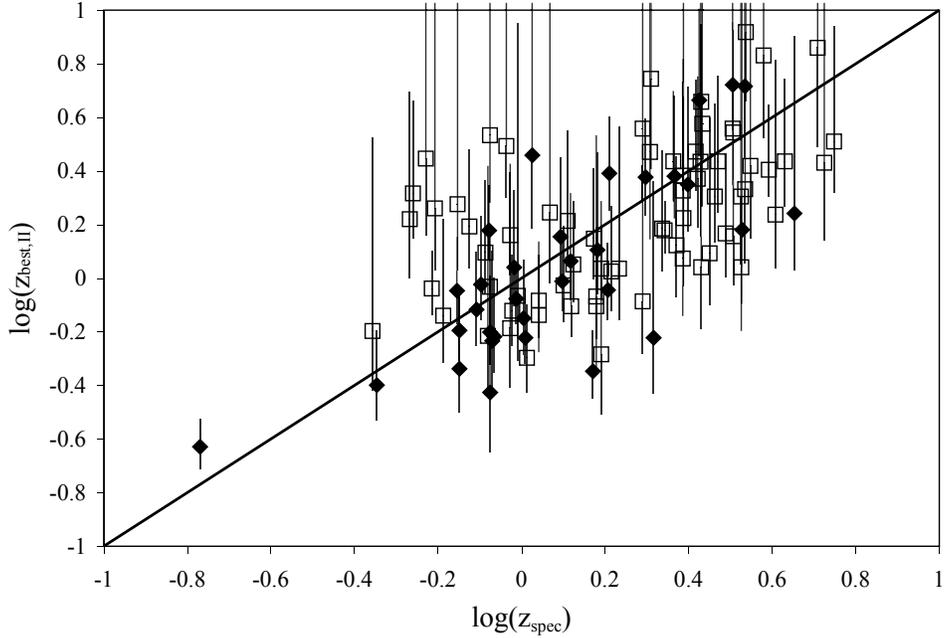} \caption{We made a comparison between our $z_{best}$ and $z_{best,II}$. $z_{best,II}$ are the calculated
redshift values when we apply \textit{Swift} luminosity relations on pre-\textit{Swift} data and pre-\textit{Swift} relations on
\textit{Swift} data. The analysis on $z_{spec,II}$ and $z_{spec}$ data shows that the average value of $log_{10}(z_{best,II}/z_{spec})$ is
-0.02, and the RMS scatter of $log_{10}(z_{best,II}/z_{spec})$ is 0.27, both of which are equal to those of our $z_{best}$ value within error
bars. From this we see that these two sets of redshifts do not differ much from each other, which also means that the pre-\textit{Swift}
luminosity relations and \textit{Swift} luminosity relations do not have significant difference. It also tells us that the effect of redshift
involved in our calculation is negligible.}

\end{figure}

\end{document}